\documentclass{article}
\usepackage[utf8]{inputenc}

\usepackage{amsmath}
\usepackage{flexisym}
\usepackage{breqn}

\usepackage{amssymb}
\usepackage{amsfonts}
\usepackage{amsthm}
\usepackage{mathtools}

\usepackage{thmtools}
\usepackage{thm-restate}

\usepackage{bbm}
\usepackage{fullpage}

\usepackage[noend]{algpseudocode}
\usepackage{algorithm}
\usepackage{algorithmicx}
\usepackage{dsfont}
\usepackage{todonotes}
\usepackage{tikz}
\usepackage{svg}
\usepackage{graphics}
\usepackage{graphicx}
\usepackage{subcaption}
\usepackage[final]{pdfpages}

\usepackage[font=small,labelfont=bf]{caption}

\declaretheorem{theorem}

\declaretheorem[numberlike=theorem]{corollary}

\usepackage{hyperref}
\hypersetup{
        colorlinks,
    linkcolor={red!50!black},
    citecolor={blue!50!black},
    urlcolor={blue!80!black}
}
\usepackage{cleveref}

\usepackage[backend=bibtex, natbib=true,url=false,maxbibnames=99,sortcites]{biblatex} 

\renewbibmacro*{doi+eprint+url}{%
    \printfield{doi}%
    \newunit\newblock%
    \iftoggle{bbx:eprint}{%
        \usebibmacro{eprint}%
    }{}%
    \newunit\newblock%
    \iffieldundef{doi}{%
        \usebibmacro{url+urldate}}%
        {}%
    }

\newcommand*\diff{\mathop{}\!\mathrm{d}}

\newcommand{\D}{\mathcal D}

\newcommand{\vis}{\textrm{vis}}
\newcommand{\visphipol}[3]{\textrm{vis}_{#3}(#1,#2)}
\newcommand{\visphi}[2]{\visphipol{#1}{#2}{}}
\newcommand{\traps}[2]{T(#1,#2)}

\newcommand{\mydef}{:=}

\newcommand{\leadstolone}{\,\includegraphics[page=16,scale=0.7]{figs.pdf}\,}

\newtoks\rowvectoks
\newcommand{\rowvec}[2]{%
  \rowvectoks={#2}\count255=#1\relax
  \advance\count255 by -1
  \rowvecnexta}
\newcommand{\rowvecnexta}{%
  \ifnum\count255>0
    \expandafter\rowvecnextb
  \else
    \begin{pmatrix}\the\rowvectoks\end{pmatrix}
  \fi}
\newcommand\rowvecnextb[1]{%
    \rowvectoks=\expandafter{\the\rowvectoks&#1}%
    \advance\count255 by -1
    \rowvecnexta}

\title{Degree of Convexity and Expected Distances in Polygons}

\author{Mikkel Abrahamsen$^\ast$ \and Viktor Fredslund-Hansen$^\ast$}
\date{October 2022}

\newcommand\blfootnote[1]{%
  \begingroup
  \renewcommand\thefootnote{}\footnote{#1}%
  \addtocounter{footnote}{-1}%
  \endgroup
}

\begin{document}

\maketitle

\blfootnote{
$^\ast$ Department of Computer Science, University of Copenhagen, \texttt{\{miab,viha\}@di.ku.dk}.
Mikkel Abrahamsen is supported by Starting Grant 1054-00032B from the Independent Research Fund Denmark under the Sapere Aude research career programme.
Viktor Fredslund-Hansen is supported by Starting Grant 7027-00050B from the
Independent Research Fund Denmark under the Sapere Aude research career programme.
Both authors are part of Basic Algorithms Research Copenhagen (BARC), supported by the VILLUM Foundation grant 16582.
}

\begin{abstract}
We present an algorithm for computing the so-called \emph{Beer-index} of a polygon $P$ in $O(n^2)$ time, where $n$ is the number of corners.
The polygon $P$ may have holes.
The Beer-index is the probability that two points chosen independently and uniformly at random in $P$ can see each other.
Given a finite set $M$ of $m$ points in a simple polygon $P$, we also show how the number of pairs in $M$ that see each other can be computed in
$O(n\log n+m^{4/3}\log^\alpha m\log n)$
time, where $\alpha<1.78$ is a constant.

We likewise study the problem of computing the expected geodesic distance between two points chosen independently and uniformly at random in a simple polygon $P$.
We show how the expected $L_1$-distance can be computed in optimal $O(n)$ time by a conceptually very simple algorithm.
We then describe an algorithm that outputs a closed-form expression for the expected $L_2$-distance in $O(n^2)$ time.
\end{abstract}

\section{Introduction}

In this paper, we consider a function defined for pairs of points in a polygon $P$, and we want to compute the expected value of the function when choosing two points uniformly and independently at random from some set $M\subset P$ (in the sequel just called ``two random points in $M$,'' for brevity).
Here, $M$ can either be a given finite set of points in $P$ or all points in $P$, i.e., $M=P$.
For a pair of points $p,q\in P$, the functions we will consider are:
\begin{itemize}
\item the indicator $[pq\subset P]\in\{0,1\}$, i.e., whether $p$ and $q$ can see each other in $P$, and
\item the length of the geodesic shortest path from $p$ to $q$ in $P$ in the $L_1$- or $L_2$-metric.
\end{itemize}
\Cref{tab:dist} summarizes our results.

\paragraph{Beer-index.}
With $M=P$ and the first function $[pq\subset P]$, we obtain the probability $B(P)$ that two random points $p,q$ in $P$ see each other,
\[
B(P)\mydef \frac 1{\vert P\vert^2}
\iint_{p\in P}\iint_{q\in P} [pq\subset P]\diff q\diff p,
\]
where $\vert \cdot\vert$ denotes the area.
Throughout the paper, our integrals are defined with respect to Lebesgue measure.
The measure $B(\cdot)$ is known as the \emph{Beer-index} after Beer~\cite{beer1973index}.

One of the well-known characterizations of a polygon $P$ being convex is that all pairs of points in $P$ can see each other.
Thus, the number $B(P)\in[0,1]$ quantifies the degree to which this holds, so it is a natural measure for the degree of convexity of $P$.
The problem of partitioning a polygon into components that are close to convex is useful in many practical settings, as such partitions provide similar benefits as convex partitions, while the number of components can be significantly smaller when the components are allowed to be slightly non-convex~\cite{DBLP:journals/cad/GhoshALL13,DBLP:journals/comgeo/LienA06}.
This motivates ways to quantify the degree of convexity and algorithms for computing these measures.
Stern~\cite{DBLP:journals/prl/Stern89} also described the measure $B(P)$, using the equivalent form
\[
B(P)= \frac 1{\vert P\vert^2} \iint_{p\in P} \left\vert \vis(p) \right\vert \diff p,
\]
where $\vis(p)=\vis_P(p)$ is the \emph{visibility polygon} of the point $p$, i.e., the region of points in $P$ that $p$ can see.
Stern suggested that $B(P)$ be estimated by measuring $\vert \vis(p) \vert$ for every grid point $p$ in $P$ from a rectangular grid.
Rote~\cite{r-dc-13} showed that $B(P)$ can be expressed as a sum of $O(n^9)$ closed-form expressions by an algorithm that uses the same amount of time.
Here, $P$ is a polygon which may have holes, and $n$ is the number of corners.
Using a bound from~\cite{DBLP:books/daglib/0086218}, Rote~\cite{rote2017note} later observed that the number is in fact bounded by $O(n^7)$.

Buchin, Kostitsyna, L{\"{o}}ffler and Silveira~\cite{DBLP:journals/algorithmica/BuchinKLS19} described another algorithm that outputs $B(P)$, again for a given polygon $P$ which may have holes.
They claimed the running time of the algorithm to be $O(n^2)$.
Unfortunately, as the algorithm is described in the journal paper~\cite{DBLP:journals/algorithmica/BuchinKLS19}, it may output $B(P)$ as a sum of $\Omega(n^4)$ closed-form expressions and likewise have a running time of $\Omega(n^4)$.
In \Cref{app:counterex}, we show that even when $P$ is a simple polygon, i.e., when $P$ has no holes, the running time of the algorithm may be $\Omega(n^4)$.
We have made the authors aware of this mistake, and they have informally described a different algorithm to us in private communication which seems to have a running time of $O(n^2\log n)$.
However, at the time of writing, the algorithm has apparently not been written down in detail.

For a given polygon $P$ with $n$ corners and possibly with holes, we describe an algorithm outputting $B(P)$ as a sum of $O(n^2)$ closed-form expressions using $O(n^2)$ time.
The algorithm performs a sort of rotational sweep around each edge $e$.
In a usual rotational sweep, a single ray is rotating around a point, but in our case, two parallel rays anchored at the endpoints of $e$ are rotating, and we keep track of everything ``seen'' from $e$ in the strip between these rays.
Computing the Beer-index $B(P)$ then boils down to computing the contribution from pairs of points in certain trapezoids with a parallel pair of rotating edges, and there are $O(n^2)$ of these trapezoids.
The algorithm is arguably simpler than the algorithm from~\cite{DBLP:journals/algorithmica/BuchinKLS19}, which relies on a transformation using geometric duality and the computation of overlays of polygons in the dual plane, whereas our algorithm deals exclusively with primitive objects in the usual primal space.

\begin{theorem}\label{thm:beer}
Given a polygon $P$ which may have holes, there is an algorithm that returns the Beer-index $B(P)$ as a sum of $O(n^2)$ closed-form expressions, each of size $O(1)$, where $n$ is the number of corners of $P$.
The algorithm uses $O(n^2)$ time and space.
\end{theorem}

\begin{table}[]
\centering
\begin{tabular}{|l|l|l|l|}
\hline
Problem & Time & Space & Reference \\
\hline
Beer-index (pol.~with holes) & $O(n^2)$ & $O(n^2)$ & \Cref{sec:beer} \\
No.~of visible pairs (points in simple pol.) & $O(n\log n+m^{4/3}\log^\alpha m\log n)$ & $O(n+m^{4/3})$ & \Cref{sec:vispairs} \\
Total $L_1$-distance (points in $\mathbb R^d$) & $O(d\cdot m\log m)$ & $O(m)$ & \Cref{sec:L1points} \\
Expected $L_1$-distance (simple pol.) & $O(n)$ & $O(n)$ & \Cref{sec:L1polygon} \\
Expected $L_2$-distance (simple pol.) & $O(n^2)$ & $O(n^2)$ & \Cref{sec:wienerL2} \\
\hline
\end{tabular}
\caption{Our results.
Here, $n$ denotes the number of corners of a polygon, $m$ denotes the number of discrete given points and $\alpha<1.78$ is a constant.}
\label{tab:dist}
\end{table}

\paragraph{Counting visible pairs of points.}
Very recently, Buchin, Custers, van der Hoog, Löffler, Popov, Roeloffzen and Staals~\cite{buchin2022segment} studied the problem of computing the number of visible pairs among $m$ points in a simple polygon $P$ with $n$ corners.
If $m$ is small compared to $n$, they suggest to use a data structure by Hershberger and Suri~\cite{DBLP:journals/jal/HershbergerS95} to test if each pair is visible in $O(\log n)$ time.
This results in an algorithm with running time $O(n+m^2\log n)$.
For the case that $m$ is large, they describe an algorithm with running time $O(n + m^{3/2+\varepsilon} \log n \log (nm))$ for any constant $\varepsilon>0$.
We present an algorithm for counting the number of visible pairs among $m$ points in time $O(n\log n+m^{4/3}\log^\alpha m\log n)$.
Our algorithm is therefore polynomially faster than the algorithms from~\cite{buchin2022segment} except if $m$ is small, i.e., $m=o(n^{2/3})$.
We furthermore believe that our algorithm serves as a simple and instructive demonstration of two important concepts in computational geometry:
(i) the use of divide-and-conquer by splitting a simple polygon along a diagonal into two parts with nearly equally many corners, and (ii) the transformation of one problem to an apparently more tractable problem using geometric duality.

\begin{theorem}\label{thm:count}
Given a simple polygon $P$ with $n$ corners and a set $M$ of $m$ points in $P$, the number of pairs of points in $M$ that can see each other can be computed in $O(n\log n+m^{4/3}\log^\alpha m\log n)$ time, where $\alpha<1.78$ is a constant.
\end{theorem}

Finding the edges of the visibility graph of a given polygon is one of the classical problems in computational geometry.
Lee~\cite{lee1978proximity} described a simple and well-known algorithm with running time $O(n^2\log n)$ already in 1979, where $n$ is the number of corners.
The algorithm works by performing a rotational sweep around all corners.
The same algorithm was described by Sharir and Schorr~\cite{DBLP:journals/siamcomp/SharirS86}, and it also appears in the book~\cite{DBLP:books/lib/BergCKO08}.
Asano, Asano, Guibas, Hershberger and Imai~\cite{DBLP:journals/algorithmica/AsanoAGHI86} and Welzl~\cite{DBLP:journals/ipl/Welzl85} gave algorithms with running time $O(n^2)$.
Hershberger~\cite{DBLP:journals/algorithmica/Hershberger89} gave an output-sensitive algorithm using $O(k)$ time to compute the visibility graph of a triangulated simple polygon, where $k$ is the number of edges in the visibility graph.
Together with Chazelle's algorithm~\cite{DBLP:journals/dcg/Chazelle91} for triangulating a simple polygon in $O(n)$ time, this yields an optimal algorithm with time $O(k)$ (note that $k=\Omega(n)$).
When the set $M$ of points in $P$ are the corners of $P$, our algorithm from \Cref{thm:count} returns the \emph{size} of the visibility graph of a simple polygon in $O(n^{4/3}\log^{\alpha+1} n)$ time.
The number $k/\binom{n}{2}\in [0,1]$ can be considered a discrete variant of the Beer-index.
Note that using the algorithms from~\cite{buchin2022segment} result in running time $\omega(n^{3/2})$.

\begin{corollary}
Given a simple polygon $P$, the number $k$ of edges in the visibility graph of $P$ can be computed in $O(n^{4/3}\log^{\alpha+1} n)$ time.
\end{corollary}

\paragraph{Expected distances.}
The problem of determining the expected distance between two random points from a given domain has a long history.
Czuber's book from 1884~\cite{czuber1884geometrische} contains calculations of the values for equilateral triangles, squares and rectangles.
Bäsel~\cite{basel2021moments} recently derived formulas for the expected distance, as well as higher moments, in regular $n$-gons for $n=3, 4, 5, 6, 8, 10, 12$.
The paper likewise contains more historical information about these problems.
In another recent paper, Bonnet, Gusakova, Thäle and Zaporozhets~\cite{BONNET2021107813} proved that the expected distance between two points in a convex body in the plane with perimeter $1$ is between $7/60$ and $1/6$, and that these bounds are tight.
They also provide bounds for higher dimensional convex bodies.
The problem of computing expected distances often appears in distribution management, transportation system analysis and airport terminal operations analysis and modeling~\cite{hsu1990expected,BANDARA199259,TOSIC19923}.

As a warmup for computing expected distances in simple polygons, we consider the conceptually simpler problem of computing the sum of $L_1$-distances between all pairs of points from a finite set $M\subset\mathbb R^d$, i.e., with no polygon involved.
We give a very simple algorithm computing the sum in $O(d\cdot m\log m)$ time, where $m=\vert M\vert$, which nicely demonstrates the technique underlying our algorithm for computing the expected $L_1$-distance in a simple polygon.

\begin{theorem}\label{thm:L1points}
Given a set $M$ of $m$ points in $\mathbb{R}^d$, the sum of pairwise distances in the $L_1$-metric between points in $M$ can be computed in $O(d\cdot m\log m)$ time.
\end{theorem}

Hsu~\cite{hsu1990expected} studied the algorithmic problems of computing the expected distance between two random points in a given polygon.
He gave a $O(n^2)$-time algorithm for the expected geodesic $L_1$-distance in a simple polygon and a $O(n^3)$-time algorithm for the expected $L_2$-distance in a \emph{convex} polygon.
In this paper, we describe a very simple algorithm for computing the expected geodesic $L_1$-distance in optimal $O(n)$ time and another algorithm outputting an expression for the $L_2$-distance in $O(n^2)$ time, both in a simple polygon.

\begin{theorem}\label{thmL1}
Given a simple polygon $P$ with $n$ corners, there is an algorithm for computing the expected geodesic $L_1$-distance between two random points in $P$ using $O(n)$ time and space.
\end{theorem}

In the case of the expected $L_2$-metric, we give a computer-assisted proof that a closed-form expression of the expected distance can be computed in $O(n^2)$ time.
The computer-assistance consists of finding closed-form expressions for certain integrals using Maple~\cite{maple}.
\Cref{app:maple} shows an example of a Maple document for calculating a closed-form expression for an integral (in this case, the integral expresses the contribution of one trapezoid to the Beer-index, but expressions for the integrals arising when computing the expected $L_2$-distance are obtained by slight variations of this code).
To the best of our knowledge, we describe the first algorithm that computes the expected $L_2$-distance between two points in a given simple polygon that may not be convex.

\begin{theorem}\label{thm:l2thm}
Given a simple polygon $P$ with $n$ corners, there is an algorithm that as an output defines $O(n^2)$ constants using closed-form expressions of size $O(1)$, where the names of previously defined constants appear in the expressions for subsequent ones.
The algorithm then outputs the expected geodesic $L_2$-distance in $P$ as a sum of a subset of these constants.
The algorithm uses $O(n^2)$ time and space.
\end{theorem}

Let us remark that if we want to output the expected distance as one single closed-form expression (i.e., not relying on predefined constants), the length of the expression would instead be $O(n^3)$.

The algorithm works by separately computing the contribution from (i) pairs of points in $P$ that see each other, and (ii) pairs where the shortest path contains a corner of $P$.
For group (i), an algorithm identical to the one for the Beer-index is used, except that we are summing over a different integral for each trapezoid.
For group (ii), we compute for each corner $u$ of $P$ the contribution from pairs of points $p,q\in P$ where the first corner of $P$ on the shortest path from $p$ to $q$ is $u$.
We show that this boils down to computing certain areas and integrals of distances from $u$ to all points on one side of a diagonal $uv$, as well as two types of integrals over points in triangles.

More prior attention has been given to the problem of computing the average distance between two vertices in a graph $G$.
This known as the \emph{Wiener-index} of $G$, and it is a fundamental measure with important applications in mathematical chemistry and appears in thousands of publications.
Note that the Wiener-index is equivalent to the sum of pairwise distances.
For more information on the problem, see the papers~\cite{DBLP:journals/talg/Cabello19, DBLP:conf/soda/GawrychowskiKMS18}.

\section{Beer-index}\label{sec:beer}
We now present an algorithm for computing the Beer-index $B(P)$ in a polygon $P$ which may have holes.
Assume without loss of generality that $P$ has area $1$ and let $n$ be the number of corners of $P$.
We furthermore assume that no three corners of $P$ are on a line.
This assumption can be avoided using a symbolic perturbation of the points~\cite{edelsbrunner1990simulation}.

\begin{figure}
\centering
\includegraphics[page=2]{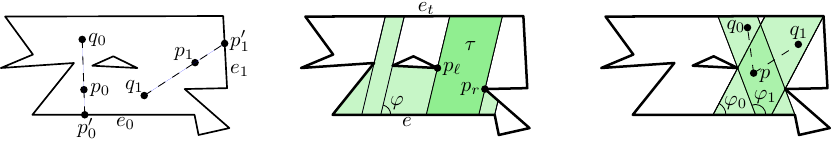}
\caption{Left: The pair $p_0,q_0$ contributes to $e_0$, since the extension $p'_0$ from $p_0$ is on $e_0$.
The pair $p_1,q_1$ contributes to $e_1$.
Middle: The polygon $\visphi e{\varphi}$ shown in green with a trapezoid $\tau$ in darker green.
The left and right pivots of $\tau$ are $p_\ell$ and $p_r$, and the top edge is $e_t$.
Right: The same trapezoid $\tau$ as in the middle, which is created at the angle $\varphi_0$ and removed at $\varphi_1$.
The pair $p,q_0$ contributes to $\tau$, but the pair $p,q_1$ does not.}
\label{fig:trapezoids}
\end{figure}

The algorithm iterates over the edges of $P$ and computes for each edge $e$ a contribution from pairs of points $p,q\in P$ that see each other and so that the extension $pp'$ of $qp$ from $p$ has the endpoint $p'$ at $e$; see \Cref{fig:trapezoids} (left).
Here, the \emph{extension} of a segment $ab\subset P$ from $b$ is the maximal segment $bb'$ on the same line as $ab$ such that $bb'\subset P$ and $ab\cap bb'=\{b\}$.
The extension from $a$ is defined analogously.
We assume without loss of generality that $e$ is horizontal with the interior of $P$ above $e$.

To compute the contribution corresponding to $e$, we perform a rotational sweep around the endpoints of $e$ with two parallel rays.
Informally, we keep track of the region seen from $e$ in the strip between the rays and in the direction of the rays.
We say that a point $q\in P$ is \emph{seen} from $e$ at an angle of $\varphi\in [0,\pi]$ if there exists $p\in e$ with $pq\subset P$ and so that the counterclockwise angle from $e$ to $pq$ is $\varphi$.
Define $\visphi e{\varphi} = \visphipol e{\varphi}{P}$, $\varphi\in [0,\pi]$, to be the region of points in $P$ seen from $e$ at an angle of $\varphi$; see \Cref{fig:trapezoids} (middle).
We keep track of the region $\visphi e{\varphi}$ as $\varphi$ increases from $0$ to $\pi$.

Let $d_\varphi=(\cos\varphi,\sin\varphi)$ be the direction corresponding to the angle $\varphi$.
For each corner $c$ of $P$ contained in $\visphi e{\varphi}$, let us draw a line through $c$ with direction $d_\varphi$.
These lines partition $\visphi e{\varphi}$ into a set of \emph{rotating trapezoids} (sometimes just called \emph{trapezoids} for brevity) which we denote as $\traps e{\varphi}$.

Each rotating trapezoid $\tau\in \traps e{\varphi}$ is in fact a map from an interval of angles $[\varphi_0,\varphi_1]$ to trapezoids contained in $P$, so that $\tau(\varphi)$ is a trapezoid in $\visphi e{\varphi}$ bounded by $e$, a pair of parallel edges with direction $d_\varphi$, and another edge of $P$, called the \emph{top} edge of $\tau$.
As the angle $\varphi$ increases, the parallel edges are thus rotating in a parallel fashion, and each is rotating around a corner of $P$.
These corners are called the \emph{left} and \emph{right pivot} of $\tau$ and the associated edges of $\tau$ are the \emph{left} and \emph{right} edges of $\tau$, respectively.

Whenever a corner of $P$ starts or stops being seen from $e$, trapezoids are created in or removed from $\traps e{\varphi}$, or both.
Consider a trapezoid $\tau$ that is created at some angle $\varphi_0$ and removed again at some angle $\varphi_1>\varphi_0$; see \Cref{fig:trapezoids} (right).
Once we get to the angle $\varphi_1$ and observe that $\tau$ is removed, we compute a contribution to $B(P)$ corresponding to $\tau$.
That is, we integrate over all pairs $p,q\in P$ where there exists $\varphi\in[\varphi_0,\varphi_1]$ such that $pq\in \tau(\varphi)$ and the oriented segment $pq$ has direction $d_\varphi$.

\paragraph{Events.}
The events of the rotational sweep are the angles $\varphi$ at which the combinatorial structure of the trapezoids of $\visphi e{\varphi}$ change.
This can happen because a corner becomes visible or invisible from $e$.
Whenever a change happens, we will compute a contribution to the Beer-index from the trapezoids that stop existing at the event.
It then follows that we compute the contribution to the Beer-index $B(P)$ of all pairs of points in $P$ that see each other, since every pair appears in a trapezoid that will eventually stop existing.
Furthermore, the pairs contributing to different trapezoids are disjoint, except for the shared boundary between the neighboring trapezoids, which is a set of measure $0$ in $P\times P$, so we do not over-count any contribution.

\begin{figure}[t]
\centering
\includegraphics[page=1,width=\textwidth]{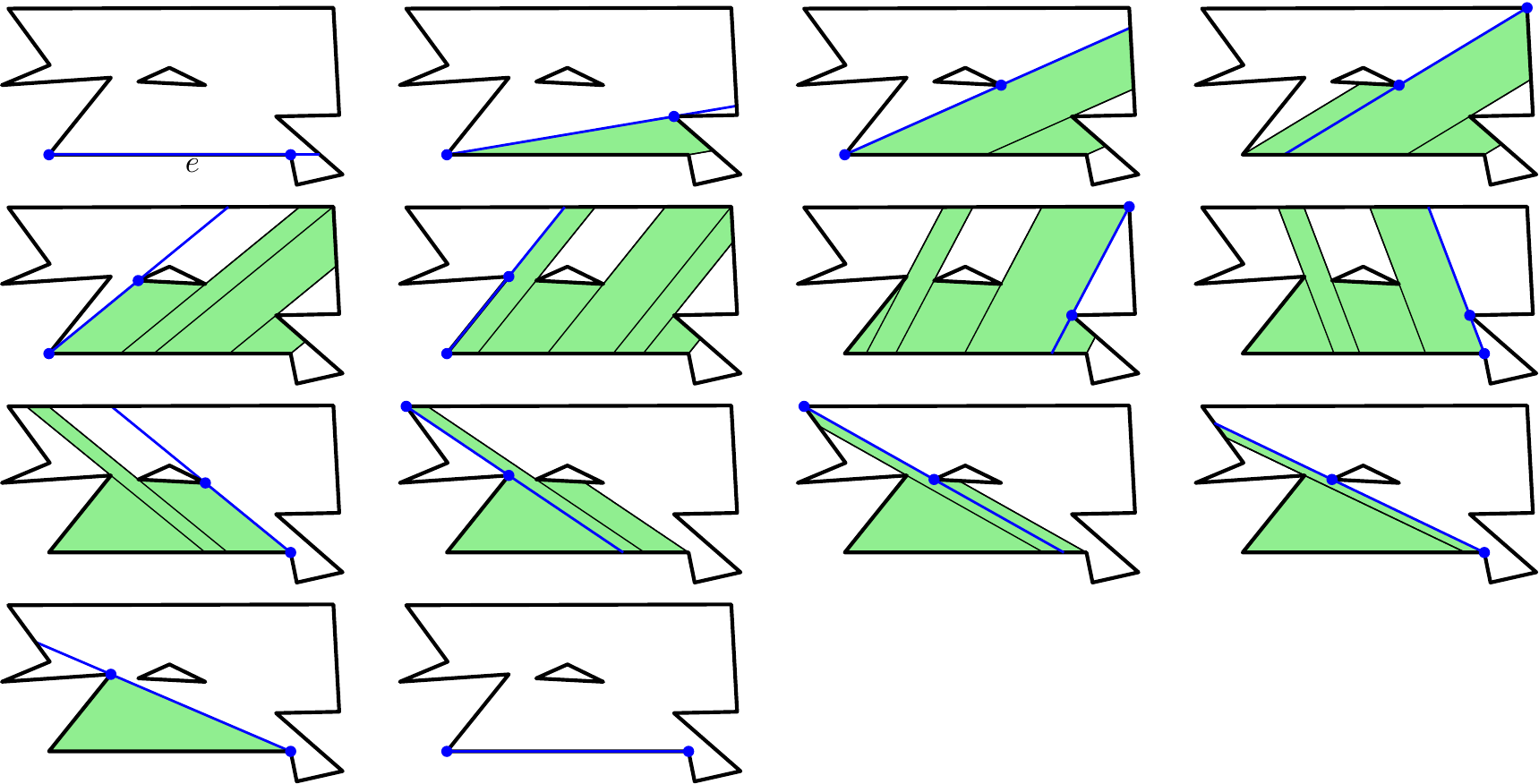}
\caption{An example with all the events that happen when scanning around the edge $e$.
The blue segments and points indicate the diagonal (and its extensions) causing the event.}
\label{fig:completeex}
\end{figure}

When a corner of $P$ becomes visible from $e$, an old trapezoid will be replaced by two new ones, and we call it an \emph{appearance} event.
Likewise, when a corner of $P$ becomes invisible from $e$, two old trapezoids will be replaced by one new, and we call it a \emph{disappearance} event.
The angles $\varphi=0$ and $\varphi=\pi$ are exceptional in that for $\varphi=0$, the first trapezoid is created and for $\varphi=\pi$, the last trapezoid is removed.
See \Cref{fig:completeex} for an example of all events occurring in one scan.

Note that when a corner $c$ becomes visible, it is because an edge $f$ stops blocking $c$ from being visible.
Likewise, when a corner $c$ stops being visible, it is because an edge $f$ starts blocking $c$ from being visible.
Therefore, appearance and disappearance events happen when $c$ and an endpoint $c'$ of $f$ are both visible at an angle of $\varphi$ from the same point on $e$.
We conclude that each event corresponds to a diagonal of $P$ whose extension intersects $e$.
(Recall that the edges of $P$ are also diagonals by definition.)
Such a diagonal will either share an endpoint with $e$ or the extension will have an endpoint at an interior point of $e$.

We first describe a version of the algorithm that handles the events in order of angle.
Here it seems unavoidable to use a priority queue to keep track of the forthcoming events.
As there are $O(n^2)$ diagonals defining the events and the queue operations take $O(\log n)$ time, we end up with a running time of $O(n^2\log n)$.
In the end of the section, we will explain how to get down to $O(n^2)$ time.
This comes at the cost that we will generally not handle the events in that order, so the faster algorithm may be conceptually more challenging to understand, although it is strictly speaking a simpler algorithm.

\paragraph{Computing diagonals.}
In order to keep track of the events, we compute for each corner $c$ the set $\D_c$ of all diagonals of $P$ with an endpoint at $c$ in order sorted by angle.
We can compute the diagonals by running a simple rotating sweep-line algorithm centered at each corner~\cite{DBLP:books/lib/BergCKO08}, which takes $O(n^2\log n)$ time in total.

\begin{figure}
\centering
\includegraphics[page=3]{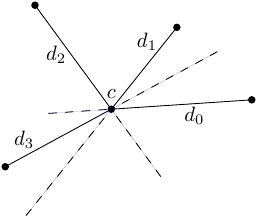}
\caption{The full segments are diagonals from the corner $c$ and the dashed segments are their extensions. The successors of $d_0,d_1,d_2,d_3$ are $d_3,d_2,d_0,d_1$, respectively.}
\label{fig:diagonals}
\end{figure}

Let $c$ be a corner of $P$ and consider a segment $cd\subset P$ and the full line $l(cd)$ containing $cd$.
The \emph{successor} of $cd$ around $c$ is the diagonal $cd_0$ we first meet when rotating $l(cd)$ counterclockwise around $c$; see \Cref{fig:diagonals}.
We will precompute the successor of each diagonal $cd\in\D_c$ as follows.
After the diagonals $\D_c$ have been computed, we can perform a rotational sweep with a full line around $c$, and every time we meet a diagonal, it is the successor of the previously encountered diagonal.
We can therefore precompute all successors of diagonals around $c$ in $O(\vert \D_c\vert)$ time.

\begin{figure}
\centering
\includegraphics[page=4]{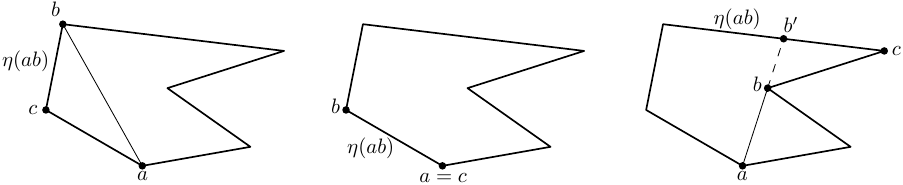}
\caption{The $\eta$-edge for three different diagonals in the same polygon showing the different cases of the definition.
Left: $c$ is to the left of $ab$. Middle: $c=a$. Right: $c$ is to the right of $ab$.}
\label{fig:etaedges}
\end{figure}

In order to find the top edges of trapezoids, we define for each oriented diagonal $ab$ of $P$ an edge $\eta(ab)$ as follows; see \Cref{fig:etaedges} for an illustration.
The edge $\eta(ab)$ has the informal description as the edge of $P$ we will see when looking ``between'' $a$ and $b$ from an angle slightly larger than that of $ab$.
Let $c$ be the corner following $b$ on the boundary $\partial P$, so that the interior of $P$ is to the left of $bc$.
If $c$ is to the left of the line containing $ab$, oriented from $a$ to $b$, or $c=a$, then we define $\eta(ab)=bc$.
Otherwise, let $bb'$ be the extension of $ab$ from $b$.
Then $\eta(ab)$ is the edge of $P$ containing $b'$.
By our general position assumption, this edge is unique.
We can easily define the $\eta$-edges while finding the diagonals with a rotational sweep around each corner.

\paragraph{Representing rotating trapezoids.}
We maintain each existing trapezoid $\tau$ as a collection of a few pointers to objects in $P$, as follows.
We use a pointer to the left and right pivot and the top edge.
For each of the two pivots, we will furthermore keep track of the successor of the current direction $d_\varphi$.
The trapezoid $\tau$ will then be removed once we reach the event corresponding to the first of these two successors.
We store the trapezoids in a priority queue $Q$ ordered by the angle of removal, so that the next trapezoid to leave $Q$ is the next to be removed.

\begin{figure}
\centering
\includegraphics[page=5]{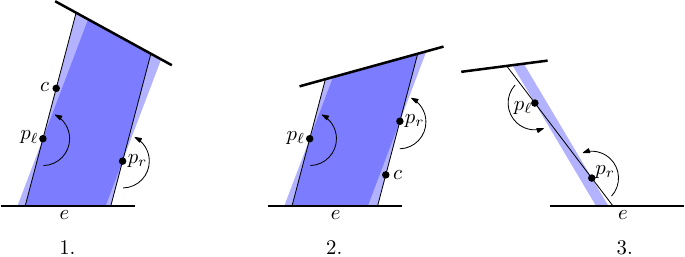}
\caption{The three ways in which a trapezoid can be removed, corresponding to cases \ref{disappear:1}--\ref{disappear:3}.}
\label{fig:events}
\end{figure}

\paragraph{Handling events.}

Consider a trapezoid $\tau$ that we dequeue from $Q$.
We will then compute the contribution to $B(P)$ corresponding to $\tau$, which will be explained later.
The trapezoid $\tau$ is removed because the left edge $e_\ell$ hits another corner than the left pivot $p_\ell$ or the right edge $e_r$ hits another corner than the right pivot $p_r$.
As the edges sweep in a counterclockwise direction, this can only happen because of one of these three cases; see~\Cref{fig:events}:
\begin{enumerate}
\item The part of $e_\ell$ above $p_\ell$ hits a corner $c$ of $P$ that becomes visible. \label{disappear:1}
\item The part of $e_r$ below $p_r$ hits a corner $c$ of $P$ and $p_r$ becomes invisible. \label{disappear:2}
\item The part of $e_\ell$ below $p_\ell$ hits $p_r$ or the part of $e_r$ above $p_r$ hits $p_\ell$. Here, $p_\ell$ becomes invisible. \label{disappear:3}
\end{enumerate}
We handle the cases in the following way.
\begin{enumerate}
\item We create two new trapezoids $\tau_0$ and $\tau_1$.
Here, $\tau_0$ is similar to $\tau$ but with $c$ as left pivot instead of $p_\ell$.
The trapezoid $\tau_1$ has $c$ as right pivot, $p_\ell$ as left pivot and $\eta(p_\ell c)$ as top edge.
\item We create a new trapezoid $\tau_0$ which is similar to $\tau$ but with $c$ as right pivot instead of $p_r$.
\item We don't create any new trapezoids.
\end{enumerate}

When a new trapezoid $\tau'$ is created, we know the diagonal $d'$ that caused $\tau'$ to be created.
Recall that we need to find the successor of $d'$ around each of the two pivots of $\tau'$.
Let us consider an appearance event as in case \ref{disappear:1}.
For the trapezoid $\tau_0$, the successor around the right pivot $p_r$ can be inherited from the removed trapezoid $\tau$, and the successor around the left pivot $c$ is the precomputed successor of the diagonal $p_\ell c$.
The successors around the pivots of the trapezoid $\tau_1$ are the precomputed successors of the diagonal $p_\ell c$.
The new trapezoid created in case \ref{disappear:2} can be handled in a similar way.
We conclude that the successors can be found in $O(1)$ time.

In order to prove that the algorithm works, we note that each trapezoid $\tau$, except the very first one, has a unique predecessor $\tau'$, which is the one whose removal causes $\tau$ to be created.
It then follows that the algorithm will create and remove each trapezoid exactly once.

\paragraph{Contribution of one rotating trapezoid.}

The perhaps most involved part of our algorithm is to compute the contribution of a single trapezoid.
Consider a trapezoid $\tau$ existing in the angle interval $[\varphi_0,\varphi_1]$.
We evaluate the contribution of $\tau$ by integrating over quadruples of coordinates $(x_0,y_0,x_1,y_1)\in P\times P$, such that the segment from $(x_0,y_0)$ to $(x_1,y_1)$ is contained in $\tau(\varphi)$ and has direction $d_\varphi$, for some $\varphi\in[\varphi_0,\varphi_1]$.

\begin{figure}
\centering
\includegraphics[page=6]{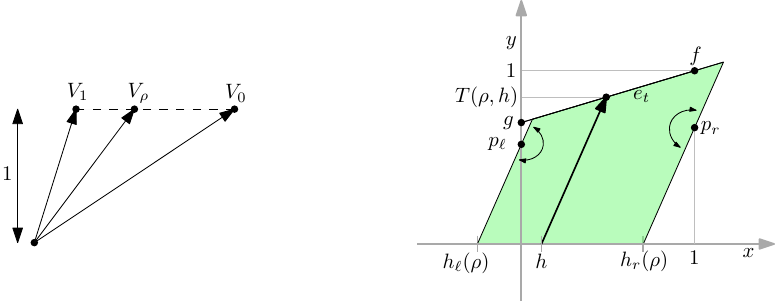}
\caption{Left: Interpolation between the defining vectors $V_0$ and $V_1$.
Right: For all points $(h,0)$ on the segment $uv$ in the trapezoid, we integrate over all pairs of points in the direction of $V_\rho$.}
\label{fig:trapezoidContribution}
\end{figure}

We will apply a sequence of transformations to $\tau$ in order to obtain an expression for the contribution of $\tau$ which is as simple as possible; see \Cref{fig:trapezoidContribution}.
In the previous description of the algorithm, we assumed without further explanation that the edge $e$ is horizontal and that the interior of $P$ is above $e$.
If $e$ does not have this property, we first apply the unique linear angle-preserving transformation $A$ that maps the endpoints $u$ and $v$ of $e$ to $(0,0)$ and $(1,0)$, respectively.
The transformation $A$ scales all distances by a factor of $s_0=1/\Vert e\Vert_2$.
Since we integrate over a set of points in $P\times P\subset\mathbb R^4$, this causes the integral to be scaled by $s_0^4$, so we should scale the final result by $1/s_0^4$.

Denote by $x(\cdot)$ and $y(\cdot)$ the $x$- and $y$-coordinates of a point or a vector, respectively.
Let the left and right pivots of $\tau$ be $p_\ell$ and $p_r$, respectively.
We use a horizontal translation to obtain that $x(p_\ell)=0$.
Suppose that $x(p_r)>0$ and define  $s_1=1/x(p_r)$.
Scaling the $x$-coordinates of all objects by $s_1$, we can assume that $x(p_r)=1$.
Since we integrate over a set of points in $P\times P\subset\mathbb R^4$ where the first and third coordinates are $x$-coordinates, this causes the integral to be scaled by $s_1^2$, so we should scale the final result by $1/s_1^2$.
If instead $x(p_r)<0$, we can obtain $x(p_r)=-1$.
In the exceptional case that $x(p_r)=0$, we already have a situation with simple coordinates and can proceed in an analogous way.

Let $f$ and $g$ be the endpoints of the top edge $e_t$ of $\tau$.
We are given $f$ and $g$ as particular points, but when computing the contribution of $\tau$, we are free to choose other points on the same line containing the edge $e_t$, i.e., we can ``slide'' $f$ and $g$ along the line containing $e_t$ and choose some points with particularly simple coordinates on the line.
If the edge $fg$ is not vertical, we can therefore assume that $x(f)=1$ and $x(g)=0$.
In the exceptional case that $fg$ is vertical, we can instead assume that $y(f)=0$ and $y(g)=1$.
In the following, we will focus on the general case where we have $x(f)=1$ and $x(g)=0$, and in the other case, we could proceed in an analogous way.
Let $s_2=1/y(f)$ and note that by scaling the $y$-coordinates of all objects by $s_2$, we obtain that $y(f)=1$.
This again causes the result of the integral to be scaled by $s_2^2$, so we should multiply the final result by $1/s_2^2$.
We then have $f=(1,1)$.

The start and end angles $\varphi_0,\varphi_1$ are defined by vectors $V_0$ and $V_1$, respectively.
Note that $y(V_0)\geq 0$ and $y(V_1)\geq 0$, as each vector points from a corner above $e$ to another corner further above $e$.
In particular, the counterclockwise angle from $V_0$ to $V_1$ is also at most $\pi$.
If the angle is at least $\pi/4$, we introduce one or two auxiliary vectors that lie in between the angular span from $V_0$ and $V_1$ so that we ``split'' the trapezoid $\tau$ into two or three trapezoids.
These have the same pivots and top edge as $\tau$, but the range of rotation of each of them is only a subset of the range of the original trapezoid $\tau$.
In the following, we therefore assume that the angle from $V_0$ to $V_1$ is less than $\pi/2$.

In the general case that $y(V_0)> 0$ and $y(V_1)> 0$, we can scale $V_0$ and $V_1$ to obtain $y(V_0)=y(V_1)=1$.
Otherwise, note that it is impossible that $y(V_0)=0$ and $y(V_1)=0$, since this would mean that a third corner of $P$ is on the line through $uv$, which is excluded by our general position assumption.
If $y(V_0)=0$ and $x(V_0)>0$, we know that $x(V_1)>0$ and $y(V_1)>0$ since the angle from $V_0$ to $V_1$ is less than $\pi/2$, so we can scale to obtain $V_0=(1,0)$ and $x(V_1)=1$.
Otherwise, we have that $y(V_1)=0$, $x(V_0)<0$ and $x(V_1)<0$, and we scale to obtain $V_0=(-1,0)$ and $x(V_1)=-1$.
In the following, we will focus on the general case where we have $y(V_0)=y(V_1)=1$.
Expressions for the contributions in the other cases can be computed in a similar way.

For $\rho\in[0,1]$, we define a corresponding interpolated vector $V_\rho=V_0\cdot (1-\rho)+V_1\cdot \rho$, which corresponds to the time where the left and right edges of $\tau$ are parallel to $V_\rho$.
In order to evaluate the contribution of $\tau$, we will use a transformation of coordinates.
Our new coordinates are quadruples $(\rho,h,p,q)$, where $h$ denotes where we ``look'' from the $x$-axis in direction $V_\rho$, and $p$ and $q$ are scalars that can be multiplied with $V_\rho$ to obtain the two points in $\tau$ we are integrating over.
The usual coordinates cooresponding to the quadruple $(\rho,h,p,q)$ are given by the transformation
\begin{align*}
F(\rho,h,p,q) & = (h+p\cdot x(V_\rho),p\cdot y(V_\rho),h+q\cdot x(V_\rho),q\cdot y(V_\rho)) \\
& = (h+p\cdot x(V_\rho),p,h+q\cdot x(V_\rho),q).
\end{align*}
We thus have the Jacobian
\[
J_F =
\begin{bmatrix}
p\cdot x(V_1-V_0) & 1 & x(V_\rho) & 0 \\
0 & 0 & 1 & 0 \\
q\cdot x(V_1-V_0) & 1 & 0 & x(V_\rho) \\
0 & 0 & 0 & 1
\end{bmatrix},
\]
with the determinant
\[
\det J_F =
(q-p)\cdot x(V_0) + (p-q)\cdot x(V_1).
\]

We now describe the bounds of the integral that we evaluate in order to compute the contribution to the Beer-index of the trapezoid $\tau$.
Let $h_\ell(\rho)$ be the $x$-coordinate of the intersection of $e$ with the line through the left pivot $p_\ell$ with direction $V_\rho$ and define $h_r(\rho)$ analogously for the right pivot $p_r$.
For a value $h\in[h_\ell(\rho),h_r(\rho)]$, let $T(\rho,h)\geq 0$ be the $y$-coordinate of the intersection of the line through $(h,0)$ with direction $V_\rho$ and the line containing $e_t$.
The contribution of $\tau$ can then be expressed as
\begin{align}
\int_0^1
\int_{h_\ell(\rho)}^{h_r(\rho)}
\int_0^{T(\rho,h)}
\int_0^{q}
\left\vert \det J_F\right\vert \diff p \diff q \diff h \diff\rho. \label{eq:int4}
\end{align}

We now show that there is a closed-form expression equal to this integral.
Note that as $q\geq p$ and $x(V_0)> x(V_1)$, we have $\det J_F> 0$, so the absolute value can be ignored.
It is straightforward to check that
\[
\int_0^{q} \det J_F \diff p = \frac{q^2(x(V_0) - x(V_1))}2.
\]

One can verify that
\[
T(\rho,h) = \frac{(h - 1)\, y(g) - h}{(x(V_0)\, (\rho-1) - x(V_1)\,\rho)\, y(g) + x(V_0)\, (1 - \rho) + x(V_1)\, \rho - 1}.
\]
Since we have the antiderivative
\[
\int \frac{q^2(x(V_0) - x(V_1))}2 \diff q = \frac{q^3\, (x(V_0) - x(V_1))}6,
\]
we get
\begin{align}
\int_0^{T(\rho,h)}
\int_0^{q}
\left\vert \det J_F\right\vert \diff p \diff q & =
\left[
\frac{q^3\, (x(V_0) - x(V_1))}6
\right]_0^{T(\rho,h)} \nonumber \\
& = \frac{((h - 1)\, y(g) - h)^3\, (x(V_0) - x(V_1))}{6\, ((x(V_0)\, (\rho-1) - x(V_1)\,\rho)\, y(g) + x(V_0)\, (1 - \rho) + x(V_1)\, \rho - 1)^3}. \label{eq:int2}
\end{align}

We note that \eqref{eq:int2} is a cubic polynomial in $h$, so it is straightforward to compute the antiderivative
\[
\int
\int_0^{T(\rho,h)}
\int_0^{q}
\left\vert \det J_F\right\vert \diff p \diff q \diff h,
\]
which will be a quartic polynomial in $h$.
Since we have
\begin{align*}
h_\ell(\rho) & = y(p_\ell)\, (x(V_0)\, (\rho-1) - x(V_1)\, \rho)\quad\quad\quad \textrm{and} \\
h_r(\rho) & = y(p_r)\, (x(V_0)\, (\rho-1) - x(V_1)\, \rho) +1,
\end{align*}
we then get that the integral
\begin{align}
\int_{h_\ell(\rho)}^{h_r(\rho)}
\int_0^{T(\rho,h)}
\int_0^{q}
\left\vert \det J_F\right\vert \diff p \diff q \diff h \label{eq:int3}
\end{align}
is a rational function in $\rho$ (note that $\rho$ appears in the denominator in \eqref{eq:int2}, so we don't get a polynomial in~$\rho$).

Now, as \eqref{eq:int3} is a rational function, it is well-known~\cite[ch.~2]{bronstein2005symbolic} that the antiderivative
\[
\int
\int_{h_\ell(\rho)}^{h_r(\rho)}
\int_0^{T(\rho,h)}
\int_0^{q}
\left\vert \det J_F\right\vert \diff p \diff q \diff h \diff\rho
\]
can be written as a function of $\rho$ involving only the usual arithmetic operations, logarithms, and arc\-tangents.
This finishes the proof that a closed-form expression exists.

Using Maple~\cite{maple}, we have computed an explicit formula for~\eqref{eq:int4} involving the five parameters $y(p_\ell)$, $y(p_r)$, $y(g)$, $x(V_0)$, $x(V_1)$.
The formula does involve logarithms, but not arctangents.
Unfortunately, the formula is too long to be written here, but it can be seen in the output from Maple in \Cref{app:maple}.

\paragraph{Running time.}
We use $O(n^2\log n)$ time to find all diagonals in sorted order around each corner of $P$.

It is not hard to construct a polygon $P$ (with holes) where we create and remove $\Omega(n^2)$ trapezoids when sweeping around a single edge $e$.
However, we claim that when scanning over all edges $e$, the total number of handled trapezoids will also be $O(n^2)$.

\begin{figure}
\centering
\includegraphics[page=15]{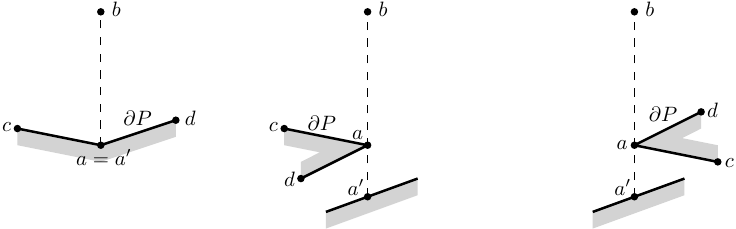}
\caption{Three cases showing when the diagonal $ab$ causes an event to happen.
The gray regions indicate the exterior of $P$.}
\label{fig:successfulldiagonals}
\end{figure}

To this end we observe that there are at most two edges such that when sweeping around these, a diagonal $ab$ can cause an event to happen.
At most two trapezoids can be created in each event, so the claim will follow.
Recall that the diagonal $ab$ will only correspond to an event when sweeping around an edge $e$ if the extension $aa'$ from $a$ intersects $e$.
Let $c$ and $d$ be the neighboring corners of $a$ on the boundary $\partial P$ such that the interior of $P$ is to the left of $ca$ and $ad$; see \Cref{fig:successfulldiagonals}.
If $c$ and $d$ are on different sides of $ab$, then it must be the case that $c$ is to the left and $d$ is to the right, since otherwise, $ab$ would not be contained in $P$.
We conclude that $a=a'$ and that $ab$ only causes an event when we sweep around $ca$ and $ad$.

Suppose on the other hand that $c$ and $d$ are on the same side of $ab$.
Then $ab$ causes an event when sweeping around the edge containing $a'$, which is unique by the general position assumption.
If $c$ and $d$ are to the left of $ab$, then $ab$ also causes an event when sweeping around $ca$, and otherwise when sweeping around $ad$.

We use $O(\log n)$ time to handle each trapezoid, which is the time it takes to dequeue from $Q$ and insert one or two trapezoids in $Q$.
Since each of the $O(n^2)$ diagonals can lead to at most two trapezoids and it takes $O(\log n)$ time to handle each trapezoid, we use $O(n^2\log n)$ time in total.

\paragraph{Getting to $O(n^2)$ time.}
Using the algorithm from~\cite{DBLP:journals/algorithmica/AsanoAGHI86}, we can compute the visibility polygon $\vis(c)$ of each corner $c$ in total time $O(n^2)$.
We then know all diagonals and can likewise compute the successors and $\eta$-edges in $O(n^2)$ total time.

We need to avoid using a priority queue, as it takes $\Omega(\log n)$ time to insert or dequeue.
We can instead use an ordinary queue (or even a stack).
We will then not handle the trapezoids in order of when they are removed, but we will still handle each trapezoid exactly once.
We then obtain \Cref{thm:beer}.

\section{Counting visible pairs}\label{sec:vispairs}

In this section, we describe an algorithm for computing the number of visible pairs among a set $M$ of $m$ points contained in a simple polygon $P$ with $n$ corners.
It is a divide-and-conquer algorithm, where we choose a diagonal $uv$ of $P$ that separates $P$ into two sub-polygons $P_1$ and $P_2$ with nearly equally many corners.
We then count the number $k$ of pairs of one point in $P_1$ and the other in $P_2$ that see each other across $uv$ and then recurse on $P_1$ and $P_2$.
We are able to compute $k$ in time $O(n+m^{4/3}\log^\alpha m)$, where $\alpha<1.78$ is a constant.
Computing the number of pairs of points in $M$ that see each other in a single triangle trivially takes $O(1)$ time once we know how many points from $M$ the triangle contains.
Let $\triangle=n-2$ be the number of triangles in a triangulation of $P$.
The recursive formula for the running time $T(\triangle,m)$ is then
\begin{align*}
T(1,m) & =O(1), \\
T(\triangle,m)&=T(\triangle_1,m_1)+T(\triangle_2,m_2)+O(\triangle+m^{4/3}\log^\alpha m), \quad \triangle>1,
\end{align*}
where $\triangle_1,\triangle_2\leq 2\triangle/3$, $\triangle_1+\triangle_2=\triangle$ and $m_1+m_2=m$.
It follows that the algorithm has running time $T(\triangle,m)=O(\triangle\log \triangle+m^{4/3}\log^\alpha m\log \triangle)=O(n\log n+m^{4/3}\log^\alpha m\log n)$.

In order to find a suitable diagonal, we first triangulate $P$, for instance using the simple algorithm from~\cite{DBLP:books/lib/BergCKO08} running in $O(n \log n)$ time.
It is well known that there exists a diagonal $uv$ in the triangulation that splits $P$ into two sub-polygons $P_1$ and $P_2$, each consisting of at most $\lceil 2\triangle/3\rceil$ triangles.
By partitioning $P_1$ and $P_2$ recursively, we obtain a balanced hierarchical decomposition of $P$ represented as a balanced binary tree $\mathcal T$.
Here, the root of $\mathcal T$ represents the diagonal $uv$, the left and right subtrees represent $P_1$ and $P_2$, respectively, and the leafs represent the individual triangles.

The triangulation of $P$ induces a dual graph $G$ which is a tree with maximum degree $3$, and the diagonal $uv$ can be chosen as incident to the triangle corresponding to the centroid of $G$.
We can therefore find $uv$ in $O(n)$ time by a simple traversal of $G$, which yields an algorithm for constructing the tree $\mathcal T$ in $O(n \log n)$ time.
By a more refined approach, $\mathcal T$ can also be computed in $O(n)$ time~\cite{DBLP:journals/algorithmica/GuibasHLST87}.

\begin{figure}
\centering
\includegraphics[page=11]{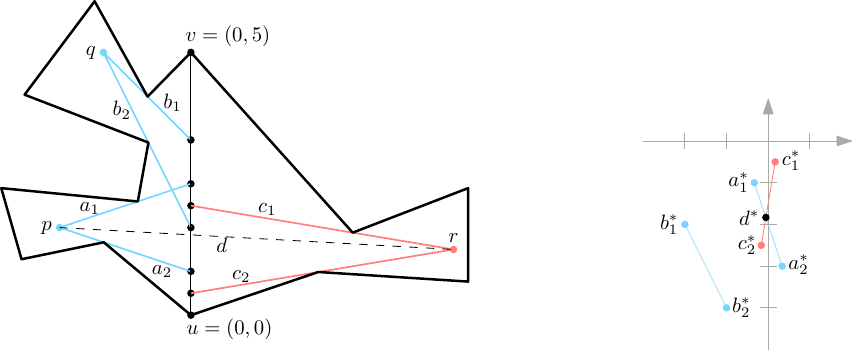}
\caption{Left: We find the visible parts of the diagonal $uv$ from the three points $p,q,r$.
Right: We map the visibility segments to the dual space and identify the segment $d=pr$ as the intersection $d^*$.}
\label{fig:counting}
\end{figure}

Our algorithm relies on the principle of geometric duality~\cite{DBLP:books/lib/BergCKO08}.
Here, a point $p=(a,b)$ is mapped to a line $p^*:y=ax-b$ and a line $\ell:y=cx+d$ is mapped to a point $\ell^*=(c,-d)$.
Assume without loss of generality that the splitting diagonal $uv$ is on the $y$-axis.
For a segment $s$, denote by $l(s)$ the full line containing $s$.
For each point $p\in M$ and every point $q\in uv$ that $p$ sees, we draw the point $l(pq)^*$ in the dual; see \Cref{fig:counting}.
Since $p$ sees an interval (or nothing) of $uv$ and all the lines $l(pq)^*$ pass through the same point $p$, these dual points form a segment $s_p$.

Our algorithm relies on the following observation:
If $p_1,p_2\in M$ are on different sides of $uv$ and the segments $s_{p_1}$ and $s_{p_2}$ intersect, then $p_1$ and $p_2$ see each other across $uv$.
To see this, note that the intersection point $r^*$ of $s_{p_1}$ and $s_{p_2}$ corresponds to a line $r$ in primal space that contains visibility segments from both $p_1$ and $p_2$ to $uv$.
Therefore, $p_1$ and $p_2$ see each other across $uv$.

When $p\in M$ is in $P_1$, we color the segment $s_p$ blue.
Otherwise, when $p$ is in $P_2$, we color $s_p$ red.
Then the number of pairs that see each other across $uv$ is exactly the number of intersections between a red and a blue segment in the dual space.
Agarwal~\cite{DBLP:journals/dcg/Agarwal90a} gave an algorithm for counting such bichromatic intersections in $O(m^{4/3}\log^\alpha m)$ time, where $\alpha<1.78$ is a constant.
The algorithm uses $O(m^{4/3})$ space.

In order to construct the segments $s_p$ efficiently, we use a data structure described by Guibas, Hershberger, Leven, Sharir and Tarjan~\cite{DBLP:journals/algorithmica/GuibasHLST87}.
Given the segment $uv$ in $P$, we can preprocess $P$ in $O(n)$ time so that given a query point $p$, we can compute the part of $uv$ that is visible from $p$ in $O(\log n)$ time.
The authors assumed that $P$ was given together with a triangulation, presumably because the paper appeared before it was known that a triangulation can be found in $O(n)$ time~\cite{DBLP:journals/dcg/Chazelle91}.

In order to handle the recursive calls, we need to find the sets $M_1=M\cap P_1$ and $M_2=M\cap P_2$.
This can be done by creating a data structure supporting ray shooting queries in $P$ in $O(\log n)$ time, as described by Hershberger and Suri~\cite{DBLP:journals/jal/HershbergerS95}.
For each point $m\in M$, we shoot a ray vertically up and check if the point where we hit the boundary is in $P_1$ or $P_2$.
The data structure takes $O(n)$ time to construct.
We therefore use
$O(n+m\log n + m^{4/3}\log^\alpha m)=O(n+m^{4/3}\log^\alpha m)$
time on preprocessing, constructing the segments $s_p$ for all $p\in M$, counting the intersections and finding the sets $M_1$ and $M_2$.
Taking the recursive calls into account, this results in an algorithm with total running time $O(n\log n+m^{4/3}\log^\alpha m\log n)$, proving \Cref{thm:count}.

\section{Expected distances in the $L_1$-metric}\label{sec:wienerL1}
\subsection{Finite set of points in $\mathbb{R}^d$}\label{sec:L1points}
Consider $m$ points $M = \left\{ p_1,p_2, \hdots, p_m \right\} \subseteq \mathbb{R}^d$, where we want to compute the sum of the $L_1$-distance between all pairs.
Our proposed method demonstrates a technique that we will also use in \Cref{sec:L1polygon} when computing the expected $L_1$-distance between a pair of random points in a polygon, so this section serves as a warmup before that algorithm is presented.
The overall idea is to compute the contribution to the total distance from each dimension separately.

\begin{figure}
\centering
\includegraphics[page=14]{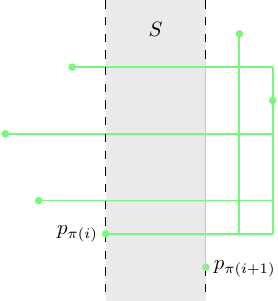}
\caption{The width of the slab $S$ is part of the difference between the $x$-coordinates of each pair of one point among $p_1,\ldots,p_{\pi(i)}$ and another among $p_{\pi(i+1)},\ldots,p_n$, and the number of such pairs is $i\cdot (m-i)$.}
\label{fig:pointsinRd}
\end{figure}

We first explain how the algorithm works in the planar case $d=2$ and then observe that it generalizes to an arbitrary dimension $d$.
When $d=2$, the algorithm works by computing the total differences of $x$- and $y$-coordinates of the points separately and then adding these numbers.
Let $\pi$ be a permutation of $\left\langle 1, 2, \ldots, m \right\rangle$ that sorts the points by $x$-coordinates, i.e., $x(p_{\pi(i)})\leq x(p_{\pi(i+1)})$ for all $1\leq i < m$. 
Let us consider the slab $S$ between the vertical lines through $x(p_{\pi(i)})$ and $x(p_{\pi(i+1)})$ and count how many pairs of points are separated by $S$; see \Cref{fig:pointsinRd}.
As there are $i$ points to the left of $S$ and $m-i$ points to the right, there are $i\cdot (m-i)$ such pairs, and the width of $S$ is part of the difference of $x$-coordinates for all these pairs.
Thus, the contribution to the total difference of $x$-coordinates of $S$ is $i\cdot (m-i) \cdot (x(p_{\pi(i+1)})-x(p_{\pi(i)}))$.
Summing over all values of $i$, we get the total difference of $x$-coordinates over all pairs of points, i.e.,
\begin{align}
\sum_{i=1}^m \sum_{j=1}^{i-1} (x(p_{\pi(i)})-x(p_{\pi(j)})) =\sum_{i=1}^{m-1} i\cdot (m-i) \cdot (x(p_{\pi(i+1)})-x(p_{\pi(i)})). \label{eq:prefix-finite-rd}
\end{align}
Note that the sum~\eqref{eq:prefix-finite-rd} can be evaluated in $O(m)$ time once the permutation $\pi$ has been found.
We can evaluate the total difference of the $y$-coordinates in an analogous way.
Hence, we can compute the total $L_1$-distance in $O(m\log m)$ time.

For points of arbitrary dimension $d$, consider a coordinate $k\in\{1,\ldots,d\}$, and let $\pi_k$ be a permutation of $\left\langle 1, 2, \ldots, m \right\rangle$ sorting the points with respect to this coordinate, i.e., we have $p_{\pi_k(i),k}\leq p_{\pi_k(i+1),k}$ for all $1\leq i < m$, where $p_{i,k}$ denotes the $k$th coordinate of $p_i$.
Using a formula analogous to~\eqref{eq:prefix-finite-rd} for each coordinate separately, we get
\begin{align}
\sum_{i=1}^m \sum_{j=1}^{i-1} ||p_i-p_j||_1 &= \sum_{i=1}^m \sum_{j=1}^{i-1} \sum_{k=1}^d |p_{i,k}-p_{j,k}| \nonumber \\
&= \sum_{k=1}^d \sum_{i=1}^m \sum_{j=1}^{i-1} |p_{i,k}-p_{j,k}| \nonumber \\
&= \sum_{k=1}^d \sum_{i=1}^{m-1} i\cdot (m-i) \cdot (p_{\pi_k(i+1),k}-p_{\pi_k(i),k}). \nonumber 
\end{align}
The above implies that the sum of all pairwise distances may be computed in time $O(d \cdot m \log m)$ by sorting the coordinates according to their $k$th coordinate and evaluating the inner sum for each $k$, which establishes \Cref{thm:L1points}.

\subsection{Expected $L_1$-distance in a polygon}\label{sec:L1polygon}
We are given a simple polygon $P$ with $n$ corners.
Let us assume for simplicity that no two corners of $P$ have the same $x$-coordinates or the same $y$-coordinates.
Note that a shortest path in $P$ in the $L_1$-metric can be realized as a union of horizontal and vertical segments.
For points $a,b\in P$, denote by $a \leadstolone b$ such a shortest path from $a$ to $b$.
More generally, when $a$ or $b$ is a compact set in $P$, we also use $a \leadstolone b$ to denote a shortest rectilinear path $P$ from a point in $a$ to a point in $b$.
Let $\vert a \leadstolone b\vert$ be the $L_1$-length of the path $a \leadstolone b$.
As in our warmup-case of points in $\mathbb R^d$ from \Cref{sec:L1points}, we find the expected distance by computing the contribution from horizontal and vertical segments separately.

Let $\delta_1(p)$ (resp.~$\delta_2(p)$) be the segments of a path $p$ that are horizontal (resp.~vertical) and define $\Delta_i(a,b) = \sum_{s \in \delta_i(a \leadstolone b)} \vert s\vert$.
Define for $A,B \subset P$ the notation
\[
D_i(A,B) = \iint_{p \in A}\iint_{q \in B} \Delta_i(p,q) \diff q\diff p.
\]
We now observe
\begin{align*}
\iint_{p \in P} \iint_{q \in P} \left|p \leadstolone q\right| \diff q\diff p  &=
\iint_{p \in P} \iint_{q \in P}  \left( \Delta_1(p, q) + \Delta_2(p, q) \right) \diff q\diff p \\
&= D_1(P,P) + D_2(P,P).
\end{align*}

\begin{figure}
\centering
\includegraphics[page=13]{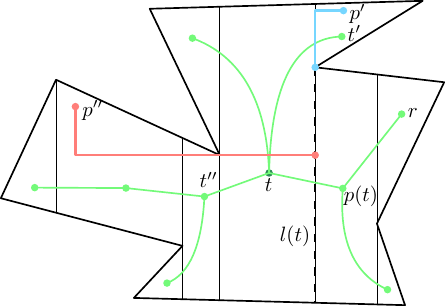}
\caption{A vertical trapezoidation of a simple polygon $P$ and the tree it induces.
Case (i) and (ii) in the computation of $L_1(t)$ are shown as shortest paths from points $p'\in P[t']$ and $p''\in P[t'']$ for children $t'$ and $t''$ of $t$, respectively, to the segment $l(t)$.}
\label{fig:trapezoidationCaseiii}
\end{figure}

We explain how to compute $D_1(P,P)$ in $O(n)$ time, and $D_2(P,P)$ can be computed in an analogous way, leading to an algorithm for computing the expected $L_1$-distance with linear running time.
We first construct a vertical trapezoidation of $P$:
For each corner $c$ of $P$, we add the maximal vertical segment contained in $P$ and containing $c$; see \Cref{fig:trapezoidationCaseiii}.
These vertical segments partition $P$ into trapezoids, each of which has a pair of vertical edges.
Some trapezoids degenerate into triangles.
The trapezoids induce a tree $T$, where the vertices are the trapezoids and two vertices are connected by an edge if the trapezoids share a vertical edge.
We choose an arbitrary root $r$ of $T$, which induce parent-child relationships among neighbouring pairs of trapezoids in $T$.
Define $p(t)$ to be the parent of $t$ and $c(t)$ to be the set of children of $t$.

For each trapezoid $t$ which is not the root, let $l(t)$ be vertical edge of $t$ that separates $t$ from the parent $p(t)$.
Furthermore, let $P[t]\subset P$ be the region consisting of $t$ and all decendants of $t$.
Define
\[
L_i(t) = \iint_{p \in P[t]} \Delta_i(p,l(t)) \diff p.
\]

We now show how we can compute the numbers $D_1(P[t],P[t])$, $L_1(t)$ and $\vert P[t]\vert$ for all trapezoids $t$ in $O(n)$ time in total.
We have then in particular computed $D_1(P,P)=D_1(P[r],P[r])$.

It is trivial to evaluate each of the numbers $D_1(P[t],P[t])$, $L_1(t)$ and $\vert P[t]\vert=\vert t\vert$ for a leaf $t$ of $T$ in $O(1)$ time, since $t$ has complexity $O(1)$, so suppose now that $t$ is not a leaf.

By our general position assumption, a trapezoid $t$ has at most four children.
Assuming that the values $D_1(P[t'],P[t'])$, $L_1(t')$ and $\vert P[t']\vert$ have been computed for each of the children $t'$ of $t$, we show how $D_1(P[t],P[t])$, $L_1(t)$ and $\vert P[t]\vert$ can be computed in $O(1)$ time.
The claim that we can compute the numbers for all $t$ in $O(n)$ time then follows.

The area $\vert P[t]\vert$ can be simply computed as
\[
\vert P[t]\vert = \vert t\vert +\sum_{t'\in c(t)} \vert P[t']\vert.
\]

We now explain how to exaluate $L_1(t)$, so suppose that $t\neq r$.
We have
\[
L_1(t) = \iint_{p \in t} \Delta_1(p,l(t)) \diff p
+\sum_{t'\in c(t)} \iint_{p \in P[t']} \Delta_1(p,l(t)) \diff p.
\]

Clearly, the first term can be evaluated in $O(1)$ time, as the trapezoid $t$ has complexity $O(1)$.
There are two cases when evaluating the integral in the sum in the second term:
(i) $t'$ and $p(t)$ are on the same side of $t$, and (ii) $t'$ and $p(t)$ are on different sides of $t$; see \Cref{fig:trapezoidationCaseiii}.
In case (i), the shortest path from a point $p\in P[t']$ to $l(t)$ first follows a shortest path to $l(t')$ and then follows a vertical segment to $l(t)$, so we have
\[
\iint_{p \in P[t']} \Delta_1(p,l(t)) \diff p =
\iint_{p \in P[t']} \Delta_1(p,l(t')) \diff p =
L_1(t').
\]

In case (ii), the shortest path first follows a shortest path from $p$ to $l(t')$ and then crosses $t$ to get to $l(t)$.
We then have
\[
\iint_{p \in P[t']} \Delta_1(p,l(t)) \diff p =
\iint_{p \in P[t']} (\Delta_1(p,l(t')) + w(t)) \diff p =
L_1(t') + \left\vert P[t']\right\vert w(t),
\]
where $w(t)$ is the width of $t$.

In order to evaluate $D_1(P[t],P[t])$, we observe
\[
D_1(P[t],P[t]) =
D_1(t,t) +
2\sum_{t'\in c(t)} D_1(P[t'],t) +
\sum_{t',t''\in c(t)} D_1(P[t'],P[t'']).
\]

Again, the first term takes $O(1)$ to evaluate as $t$ has complexity $O(1)$.
To evaluate the sum in the second term, we observe
\begin{align*}
D_1(P[t'],t) & =
\iint_{p' \in P[t']} \iint_{p \in t} \Delta_1(p',p) \diff p\diff p' \\
& = \iint_{p' \in P[t']} \iint_{p \in t} (\Delta_1(p',l(t')) + \Delta_1(l(t'),p)) \diff p\diff p' \\
& = \left\vert t\right\vert L_1(t') + \left\vert P[t']\right\vert \iint_{p\in t} \Delta_1(l(t'),p)\diff p.
\end{align*}
Again, the integral in the final expression can be evaluated in $O(1)$ time, so the same holds for $D_1(P[t'],t)$.

Consider now a term of the form $D_1(P[t'],P[t''])$.
If $t'=t''$, the number $D_1(P[t'],P[t''])$ has already been computed, so suppose $t'\neq t''$.
We have two cases:
(i) $t'$ and $t''$ are on the same side of $t$ and (ii) $t'$ and $t''$ are on different sides.
In case (i), we get
\begin{align*}
D_1(P[t'],P[t'']) & =
\iint_{p' \in P[t']}
\iint_{p'' \in P[t'']} \Delta_1(p',p'') \diff p''\diff p' \\
& = \iint_{p' \in P[t']}
\iint_{p'' \in P[t'']} (\Delta_1(p',l(t')) + \Delta_1(l(t''),p'')) \diff p''\diff p' \\
& = \left\vert P[t']\right \vert L_1(t'') + \left \vert P[t'']\right \vert L_1(t').
\end{align*}

In case (ii), we get
\begin{align*}
D_1(P[t'],P[t'']) & = \iint_{p' \in P[t']}
\iint_{p'' \in P[t'']} \Delta_1(p',p'') \diff p''\diff p' \\
& = \iint_{p' \in P[t']}
\iint_{p'' \in P[t'']} (\Delta_1(p',l(t')) + w(t) + \Delta_1(l(t''),p'')) \diff p''\diff p' \\
& = \left\vert P[t']\right\vert  L_1(t'') + \left \vert P[t'']\right\vert L_1(t') + \left\vert P[t']\right\vert \left\vert P[t'']\right\vert w(t).
\end{align*}

Since a trapezoidation of $P$ can be computed in $O(n)$ time using Chazelle's algorithm~\cite{DBLP:journals/dcg/Chazelle91}, we get \Cref{thmL1}.
Instead of Chazelle's algorithm, which is known for being complicated, we can also use a simpler sweep-line algorithm~\cite{DBLP:books/lib/BergCKO08} and obtain an algorithm running in time $O(n\log n)$.

\section{Expected distances in the $L_2$-metric}\label{sec:wienerL2}
We now show how to compute the expected geodesic $L_2$-distance between two random points in a simple polygon $P$.
For $p,q\in P$, we denote by $p \leadsto q$ the unique shortest path in $P$ from $p$ to $q$ in the $L_2$-metric, and we denote by $\Vert p \leadsto q\Vert$ the length of this path.
We thus want to compute
\[
\mathbb E(\Vert p \leadsto q\Vert) = \frac{1}{\vert P\vert^2} \iint_{p \in P} \iint_{q\in P} \left\Vert p \leadsto q\right\Vert \diff q\diff p.
\]

\begin{figure}
\centering
\includegraphics[page=17]{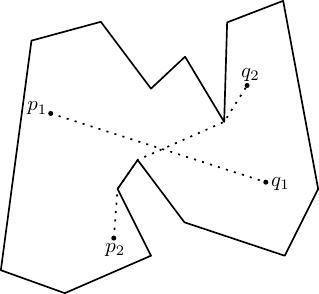}
\caption{A polygon $P$ with $(p_1,q_1)\in A$ and $(p_2,q_2)\in B$.}
\label{fig:setsAandB}
\end{figure}

We split this integral into two parts; see \Cref{fig:setsAandB}.
Let $A\subset P\times P$ be the pairs $(p,q)$ such that $p \leadsto q=pq$, i.e., $p$ and $q$ see each other in $P$.
Let $B\subset P\times P$ be the pairs $(p,q)$ such that $p \leadsto q$ contains one or more corners of $P$.
Then $A\cup B=P\times P$.
Furthermore, the intersection $A\cap B$ contains the pairs $(p,q)$ such that $pq\subset P$ and $pq$ contains a corner of $P$.
This set has measure $0$ in $P\times P$, so we have
\[
\iint_{p \in P} \iint_{q\in P} \left\Vert p \leadsto q\right\Vert \diff q\diff p =
\iiiint_{(p,q)\in A} \left\Vert p q\right\Vert \diff (p,q) +
\iiiint_{(p,q)\in B} \left\Vert p \leadsto q\right\Vert \diff (p,q).
\]
In the following subsections, we describe how we compute the integral over $A$ and $B$, respectively.
In each subsection, we explain how to output $O(n^2)$ constants defined as closed-form expressions of size $O(1)$.
As described in \Cref{thm:l2thm}, some of these will refer to previously defined constants, and in the end, the integral over both $A$ and $B$ can be expressed as a sum of a subset of these $O(n^2)$ constants, and the theorem follows.

\subsection{Shortest paths that are segments}
Let us first consider the computation of the integral over $A$.
Note that when computing the Beer-index as described in \Cref{sec:beer}, we were also integrating over the set $A$ of pairs of points that see each other.
We will therefore use a similar algorithm here that evaluates the contribution from each rotating trapezoid.
The only difference is that instead of computing the volume in $\mathbb R^4$ corresponding to each trapezoid, we will now integrate the distance function.
In other words, instead of the integral \eqref{eq:int4}, we end up with
\begin{align}
& \int_0^1
\int_{h_\ell(\rho)}^{h_r(\rho)}
\int_0^{T(\rho,h)}
\int_0^{q}
\left\Vert q\cdot V_\rho - p\cdot V_\rho\right\Vert
\left\vert \det J_F\right\vert \diff p \diff q \diff h \diff\rho = \nonumber\\
& \int_0^1
\int_{h_\ell(\rho)}^{h_r(\rho)}
\int_0^{T(\rho,h)}
\int_0^{q}
\left\Vert V_\rho\right\Vert
\left( q-p\right)
\left\vert \det J_F\right\vert \diff p \diff q \diff h \diff\rho. \label{eq:intWienerTrapez}
\end{align}

We would again like to transform the trapezoid to make it simpler, but here we need to be more careful than we were in the case of the Beer-index.
In particular, we cannot scale the $x$- and $y$-coordinates independently, since such a scaling does not just scale the Euclidean distance between two points.
We can however scale $x$- and $y$-coordinates simultaneously by some factor $s>0$.
By a transformation of coordinates, we have for any set $X\subset P\times P$ that
\begin{align*}
\iiiint_{(p,q)\in X} \left\Vert p - q\right\Vert \diff (p,q) & =
\iiiint_{(p,q)\in sX} \left\Vert p/s - q/s\right\Vert \left\vert\det (I_4/s)\right\vert\diff (p,q) \\
& =
\frac 1{s^5} \iiiint_{(p,q)\in sX} \left\Vert p - q\right\Vert \diff (p,q),
\end{align*}
where $I_4$ denotes the $4\times 4$ identity matrix.
Using this, we can obtain the following properties by scaling the final result suitably:
\begin{enumerate}
\item By a rotation and scaling, the edge $e$ is contained in the $x$-axis and the interior of $P$ is above $e$.
\item By a horizontal translation, we have $x(p_\ell)=0$, where $p_\ell$ is the left pivot.
\item By a scaling, we have $x(p_r)=1$, where $p_r$ is the right pivot.\label{item:wiener3}
\item By sliding the points $f$ and $g$ defining the top edge, we have $x(f)=1$ and $x(g)=0$.\label{item:wiener4}
\item By scaling the vectors $V_0$ and $V_1$, we have $y(V_0)=y(V_1)=1$.\label{item:wiener5}
\end{enumerate}

Again, it is not always possible to obtain properties \ref{item:wiener3}--\ref{item:wiener5} due to some special cases, but when it is not possible, we can get similar properties and proceed in an analogous way.

We obtain a rotating trapezoid depending on the six parameters $y(p_\ell),y(p_r),y(f),y(g),x(V_0),x(V_1)$.
Note that in the case of the Beer-index, we had $y(f)=1$, but we do not see how to obtain that here (while also maintaining the other properties).
We can now rewrite the integral \eqref{eq:intWienerTrapez} as
\[
\int_0^1
\int_{h_\ell(\rho)}^{h_r(\rho)}
\int_0^{T(\rho,h)}
\int_0^{q}
\sqrt{1+((1-\rho)\, x(V_0)+\rho\,  x(V_1))^2}
\left( q-p\right) 
\left\vert \det J_F\right\vert \diff p \diff q \diff h \diff\rho.
\]

Maple~\cite{maple} is able to compute a closed-form expression of this integral, involving only elementary functions.
Unfortunately, when written as clear text and using 3-letter names for the six parameters mentioned above, the formula has a length of more than 12 million characters, so it is hardly of any practical use.
We conclude that using $O(n^2)$ time, we can output $O(n^2)$ closed-form expressions of size $O(1)$ whose sum equals the integral over $A$.

\subsection{Shortest paths containing corners}
\label{sec:L2Corners}

\begin{figure}
\centering
\includegraphics[page=18]{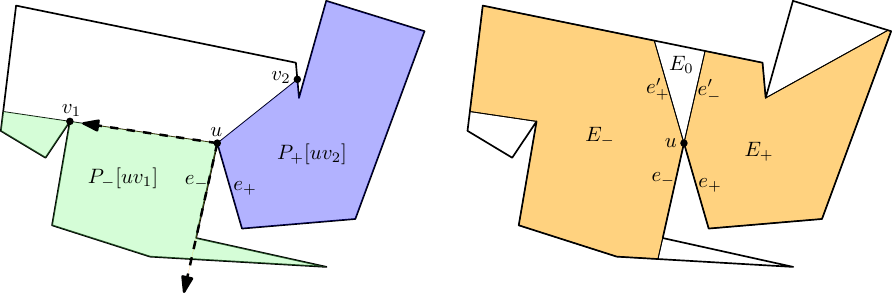}
\caption{Left: The regions $P_-[uv_1]$ and $P_+[uv_2]$ for chords $uv_1$ and $uv_2$.
The dashed arrows bound the wedge from $uv_1$ counterclockwise to $e_-$.
Right: The sets $E_-,E_0,E_+$ for a corner~$u$.}
\label{fig:setsEandE}
\end{figure}

For a corner $u$ of $P$, let $K_u\subset B$ be pairs of points $(p,q)$ such that the first corner of $P$ on the shortest path $p\leadsto q$ from $p$ is $u$.
We then define
\[
M_u=\iiiint_{(p,q)\in K_u} \left\Vert p \leadsto q\right\Vert \diff (p,q)
\]
and thus have
\begin{align}
\iiiint_{(p,q)\in B} \left\Vert p \leadsto q\right\Vert \diff (p,q) =
\sum_{u\in\mathcal C} M_u, \label{eq:wienerCornerInt}
\end{align}
where $\mathcal C$ is the set of corners of $P$.
We will compute the contribution to the integral over $B$ for each set $K_u$.
To this end, we introduce some regions and integrals defined by diagonals of $P$.
Consider a chord $uv$ of $P$ where $u$ is a corner.
Let $e_-,e_+$ be the edges of $P$ preceding and succeeding $u$ in counterclockwise order, respectively; see \Cref{fig:setsEandE} (left).
We define $P_-[uv]\subset P$ to be the points $q$ in $P$ such that the shortest path from $u$ to $q$ starts with a segment in the closed wedge from $uv$ counterclockwise to $e_-$.
Similarly, $P_+[uv]\subset P$ is the set of points $q$ in $P$ such that the shortest path from $u$ to $q$ starts with a segment in the closed wedge from $uv$ clockwise to $e_+$.
We also include $u$ in $P_-[uv]$ and $P_+[uv]$ to get closed sets.
We note that $P$ can be written as an interior-disjoint union $P=P_-[uv]\cup P_+[uv]$.
If $uv$ is a diagonal, we say that $P_-[uv]$ and $P_+[uv]$ are \emph{diagonal regions}.

For $\pm\in\{-,+\}$, we now define
\[
L_\pm[uv] = \iint_{q\in P_\pm[uv]} \left\Vert u \leadsto q\right\Vert \diff q.
\]
We will be explain in \Cref{sec:computeLints} how the areas $\vert P_\pm[uv] \vert$ and values $L_\pm[uv]$ can be computed for all diagonals $uv$.
Here we will assume that it has already been done, and we will then see how the integral $M_u$ can be computed.
First, observe that if $u$ is a convex corner, then $K_u= \vis(u)\times \{u\}$, where $\vis(u)$ is the visibility polygon of $u$.
This set has measure $0$ in $P\times P$, so clearly, $M_u=0$ in this case.
Assume therefore that $u$ is a concave corner.
Note that $K_u$ is a subset of $\vis(u)\times P$.
We now partition $\vis(u)$ into three parts as follows.
Let $e'_-$ and $e'_+$ be the extensions of $e_-$ and $e_+$ from $u$, respectively.
Then $e'_-$ and $e'_+$ partition $\vis(u)$ into three parts $E_-,E_0,E_+$, such that $e_-$ is on the boundary of $E_-$, $e_+$ is on the boundary of $E_+$, and $E_0$ is in between $e'_-$ and $e'_+$; see \Cref{fig:setsEandE} (right).
We can now partition $K_u$ into the three sets $K_u\cap (E_-\times P)$, $K_u\cap (E_0\times P)$ and $K_u\cap (E_+\times P)$.
We note that $K_u\cap (E_0\times P)=E_0\times\{u\}$, which has measure $0$, so we can express $M_u$ as
\[
M_u =
\iiiint_{(p,q)\in K_u\cap (E_-\times P)} \left\Vert p \leadsto q\right\Vert \diff (p,q) + \iiiint_{(p,q)\in K_u\cap (E_+\times P)} \left\Vert p \leadsto q\right\Vert \diff (p,q).
\]
We describe how to evaluate the integral over $K_u\cap (E_+\times P)$, and the one over $K_u\cap (E_-\times P)$ can be handled in an analogous way.

\begin{figure}
\centering
\includegraphics[page=19]{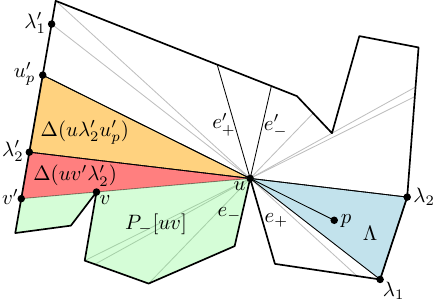}
\caption{A triangle $\Lambda$ in $E_+$.
The shortest paths from $p$ that pass through $u$ have endpoints in $\Delta(u\lambda'_2 u'_p)$, $\Delta(uv'\lambda'_2)$ or $P_-[uv]$.}
\label{fig:lambda}
\end{figure}

The diagonals $uv$ and their extensions from $u$ partition $\vis(u)$ into triangles; see \Cref{fig:lambda}.
For each triangle $\Lambda$ in $E_+$, we will compute the contribution from paths from $p\in\Lambda$ to points $q$ such that the first corner of $P$ on $p\leadsto q$ is $u$,
\[
M_u^\Lambda=\iiiint_{(p,q)\in K_u\cap (\Lambda\times P)} \left\Vert p \leadsto q\right\Vert \diff (p,q).
\]
Taking the sum of this integral over all such triangles $\Lambda$ in $E_+$ will then give us the integral over $K_u\cap (E_+\times P)$.

For a point $p\in \vis (u)$, let the extension of $pu$ from $u$ be $uu'_p$.
We can then write
\[
M_u^\Lambda=\iint_{p\in \Lambda} \iint_{q\in P_-[uu'_p]} \left\Vert p \leadsto q\right\Vert \diff q\diff p
= \iint_{p\in \Lambda} \iint_{q\in P_-[uu'_p]} \left\Vert pu\right\Vert+\left\Vert u \leadsto q\right\Vert \diff q\diff p.
\]
Let $uv$ be the diagonal which is the next after $uu'_p$ in counterclockwise direction around $u$ and let $vv'$ be the extension of $uv$ from $v$ (where we may have $v'=v$).
Note that this diagonal $uv$ does not depend on the choice of $p\in \Lambda$, so we can partition $P_-[uu'_p]$ as $P_-[uu'_p]=P_-[uv]\cup \Delta(uv'u'_p)$, and we get
\begin{align}
M_u^\Lambda & =\iint_{p\in \Lambda}
\left(
\iint_{q\in P_-[uv]} \left\Vert pu\right\Vert+\left\Vert u \leadsto q\right\Vert \diff q
+\iint_{q\in \Delta(uv'u'_p)} \left\Vert pu\right\Vert+\left\Vert uq\right\Vert \diff q\right)\diff p \nonumber \\
& = \left\vert P_-[uv]\right\vert \iint_{p\in \Lambda} \left\Vert pu\right\Vert \diff p 
+ \left\vert \Lambda\right\vert L_-[uv] 
+ \iint_{p\in \Lambda} \iint_{q\in \Delta(uv'u'_p)} \left\Vert pu\right\Vert+\left\Vert uq\right\Vert \diff q\diff p. \label{eq:int10a}
\end{align}

Let the corners of $\Lambda$ be $u,\lambda_1,\lambda_2$ in counterclockwise order.
Let $u\lambda'_1$ and $u\lambda'_2$ be the extensions of $\lambda_1u$ and $\lambda_2u$ from $u$, so that we have an interior-disjoint union $\Delta(uv'u'_p)=\Delta(uv'\lambda'_2)\cup \Delta(u\lambda'_2u'_p)$.
We can rewrite the last integral as
\[
\iint_{p\in \Lambda} \iint_{q\in \Delta(uv'\lambda'_2)} \left\Vert pu\right\Vert+\left\Vert uq\right\Vert \diff q\diff p +
\iint_{p\in \Lambda} \iint_{q\in \Delta(u\lambda'_2u'_p)} \left\Vert pu\right\Vert+\left\Vert uq\right\Vert \diff q\diff p.
\]
 Here, the first part equals
\begin{align}
\left\vert \Delta(uv'\lambda'_2)\right\vert \iint_{p\in \Lambda} \left\Vert pu\right\Vert \diff p
+
\left\vert \Lambda \right\vert \iint_{q\in \Delta(uv'\lambda'_2)} \left\Vert uq\right\Vert \diff q, \label{eq:int10b}
\end{align}
and the second equals
\begin{align}
\iint_{p\in \Lambda} \left\Vert pu\right\Vert \left\vert \Delta(u\lambda'_2u'_p)\right\vert \diff p
+\iint_{p\in \Lambda} \iint_{q\in \Delta(u\lambda'_2u'_p)} \left\Vert uq\right\Vert \diff q\diff p. \label{eq:int12}
\end{align}

By changing order of integration, we obtain
\begin{align}
& \iint_{p\in \Lambda} \iint_{q\in \Delta(u\lambda'_2u'_p)} \left\Vert uq\right\Vert \diff q\diff p= \nonumber\\
& \iint_{q\in \Delta(u\lambda'_2\lambda'_1)} \iint_{p\in \Delta(u\lambda_1 u'_q)} \left\Vert uq\right\Vert \diff q\diff p= \nonumber\\
& \iint_{q\in \Delta(u\lambda'_2\lambda'_1)}  \left\Vert uq\right\Vert \left\vert \Delta(u\lambda_1 u'_q)\right\vert \diff q. \label{eq:int13}
\end{align}

It therefore remains to explain how we compute each of the integrals
\begin{align}
&\iint_{p\in \Lambda} \left\Vert pu\right\Vert \diff p, & \textrm{[in \eqref{eq:int10a} and \eqref{eq:int10b}]} \label{eq:triint1} \\
&\iint_{q\in \Delta(uv'\lambda'_2)} \left\Vert uq\right\Vert \diff q, & \textrm{[in \eqref{eq:int10b}]} \label{eq:triint1a}\\
&\iint_{p\in \Lambda} \left\Vert pu\right\Vert \left\vert \Delta(uv'u'_p)\right\vert \diff p, & \textrm{[in \eqref{eq:int12}]} \label{eq:triint2}\\
&\iint_{q\in \Delta(u\lambda'_2\lambda'_1)} \left\Vert uq\right\Vert \left\vert \Delta(u\lambda_1 u'_q)\right\vert \diff q.  & \textrm{[in \eqref{eq:int13}]} \label{eq:triint3}
\end{align}

For three points $a,b,c$, define the integral $\Xi(abc)$ over the triangle $\Delta(abc)$ as
\[
\Xi(abc)=
\iint_{q\in \Delta(abc)} \left\Vert aq\right\Vert \diff q.
\]
We then note that \eqref{eq:triint1} is just $\Xi(u\lambda_1\lambda_2)$ and \eqref{eq:triint1a} is $\Xi(uv'\lambda'_2)$.
Furthermore, note that \eqref{eq:triint2} and \eqref{eq:triint3} have the same form, so that a closed-form expression for one automatically gives one for the other as well.
In \Cref{sec:triangleintegrals}, we show how the integrals \eqref{eq:triint1}--\eqref{eq:triint3} can be computed using Maple.

Given that we know all diagonals $uv$ from $u$, it takes $O(n)$ time to compute the contribution $M_u$ by iterating over all triangles $\Lambda$ in $E_+$ and $E_-$.
We therefore use $O(n^2)$ time in total on outputting closed-form expressions, each of size $O(1)$, whose sum equals the integral over $B$.

\subsection{Computing the values $\vert P_\pm[uv]\vert$ and $L_\pm(uv)$}\label{sec:computeLints}

\begin{figure}
\centering
\includegraphics[page=20]{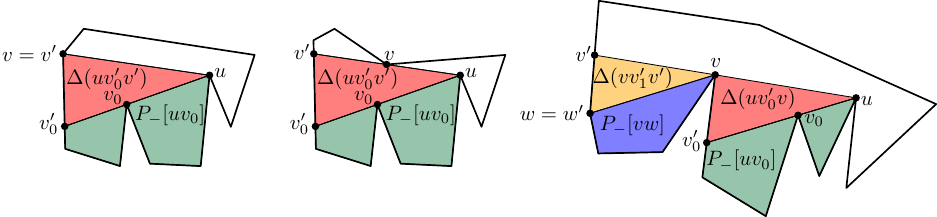}
\caption{The figures show how a diagonal region $P_-[uv]$ can be described as interior-disjoint unions of smaller diagonal regions and triangles.
This depends on which regions $P_-[uv]$ and $P_+[uv]$ contain the edges incident at $v$.
The region $P_-[uv]$ is colored and $P_+[uv]$ is white.
Left: One edge is in $P_+[uv]$.
Middle: Both edges are in $P_+[uv]$.
Right: Both edges are in $P_-[uv]$.}
\label{fig:diagonalregions}
\end{figure}

We explain how the values $\vert P_-[uv]\vert$ and $L_-(uv)$ are computed; the values $\vert P_+[uv]\vert$ and $L_+(uv)$ are computed in an analogous way.
We first observe that each diagonal region $P_-[uv]$ can be described as an interior-disjoint union of at most two smaller diagonal regions and at most two triangles.
Let $vv'$ be the extension of $uv$ from $v$.
We now have two cases.
\begin{enumerate}
\item One or both of the edges of $P$ incident at $v$ are in $P_+[uv]$; see \Cref{fig:diagonalregions} (left and middle).
Let $uv_0$ be the diagonal after $uv$ in counterclockwise order around $u$, and let $v_0v_0'$ be the extension of $uv_0$ from $v_0$.
Then
\begin{align}
P_-[uv]=P_-[uv_0]\cup \Delta(uv'_0v').\label{eq:union1}
\end{align}

\item Both edges of $P$ incident at $v$ are in $P_-[uv]$; see \Cref{fig:diagonalregions} (right).
Let $uv_0$ be the diagonal after $uv$ in counterclockwise order around $u$, and let $v_0v_0'$ be the extension of $uv_0$ from $v_0$.
Let $vw$ be the diagonal after $vv'$ in counterclockwise order around $v$, and let $ww'$ be the extension of $vw$ from $w$.
Then
\begin{align}
P_-[uv]=P_-[uv_0]\cup \Delta(uv'_0v)\cup P_-[vw]\cup \Delta(vw'v').\label{eq:union2}
\end{align}
\end{enumerate}

These formulas naturally lead to a recursive approach for computing the values $\vert P_-[uv]\vert$.
Recall that we can compute all diagonals of $P$ in $O(n^2)$ time.
For each diagonal $uv$ we then compute the area $\vert P_-[uv]\vert$ using \eqref{eq:union1} or \eqref{eq:union2}.
When a value for a smaller region is needed which we have not yet computed, we do it recursively.
We use $O(1)$ time per diagonal $uv$ to compute the area $\vert P_-[uv]\vert$.

Using the notation $\Xi(\cdot)$ introduced in \Cref{sec:L2Corners} for integrals of distances over triangles, we get the following two formulas corresponding to~\eqref{eq:union1} and~\eqref{eq:union2}.
\begin{align}
L_-[uv] & =L_-[uv_0] + \Xi(uv'_0v), \label{eq:computeLint1} \\
L_-[uv] & =
L_-[uv_0] + \Xi(uv'_0v)+
L_-[vw]+
\Xi(vw'v')+
(\left\vert P_-[vw]\right\vert +
\left\vert \Delta(vw'v')\right\vert) \left\Vert uv\right\Vert.\label{eq:computeLint2}
\end{align}

We can therefore likewise compute all the values $L_-[uv]$ recursively.
In total, closed-form expressions for all the values $\vert P_\pm[uv]\vert$ and $L_\pm[uv]$ can be returned in $O(n^2)$ time.

\begin{figure}
\centering
\includegraphics{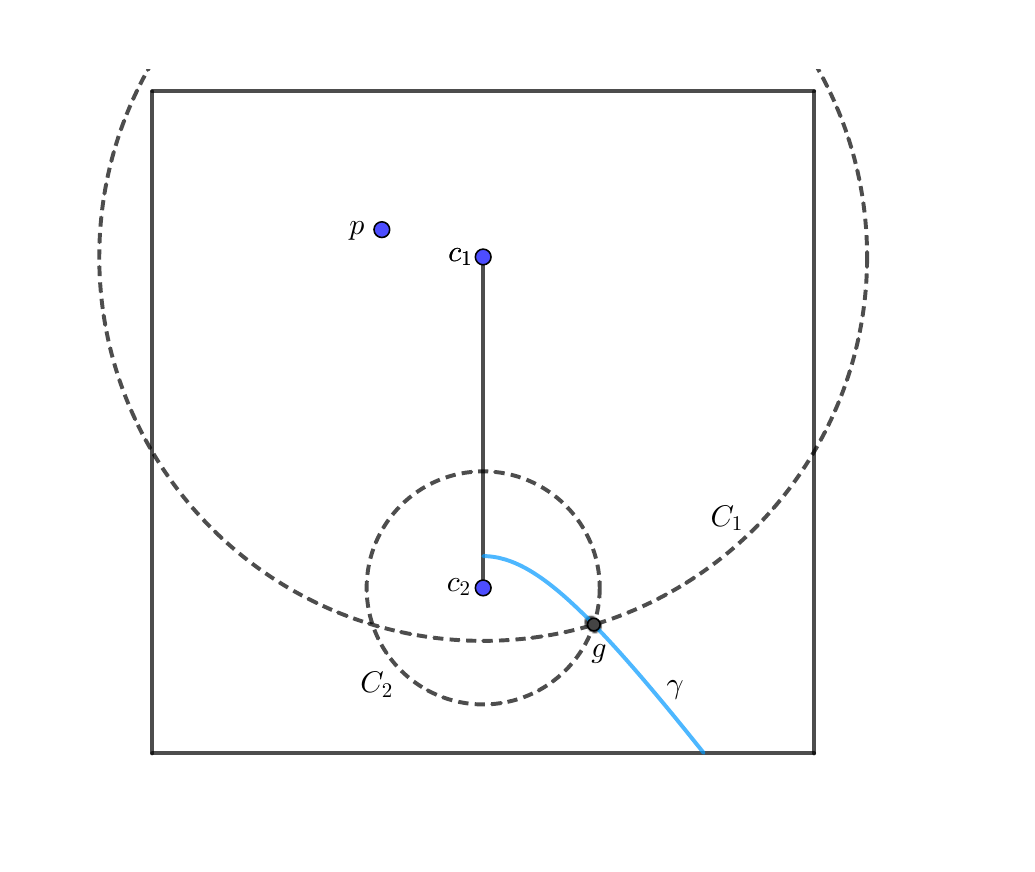}
\caption{A square polygon $P$ where the segment $c_1c_2$ is a (degenerate) hole.
The curve $\gamma$ is the locus of points $g$ where the paths from $p$ to $g$ above $c_1$ and below $c_2$ are equally long.
Each such point is an intersection of circles $C_1$ and $C_2$ centered at $c_1$ and $c_2$, where the radius of $C_1$ is $\Vert pc_2\Vert - \Vert pc_1\Vert$ larger than the radius of $C_1$.
Therefore, $\gamma$ is an arc on a hyperbola.}
\label{fig:hyperbola}
\end{figure}

\section{Concluding remarks}

It is interesting whether the expected distance between two points in a polygon with holes can also be expressed in closed form.
This seems much harder.
Even the expected distance from a fixed point $p$ to another random point is complicated to compute in the presence of holes; see \Cref{fig:hyperbola}.
Whether the shortest path from $p$ to another point $q$ goes above or below a single hole sometimes depends on whether $q$ is above or below an arc $\gamma$ on a certain hyperbola which depends on $p$.

Another interesting question is whether our $O(n^2)$-time algorithms for the Beer-index and the expected $L_2$-distance are conditionally optimal under some reasonable hypothesis.
We could not find faster algorithms, even for simple polygons when computing the Beer-index and convex polygons when computing the expected distance.

\printbibliography

\appendix
\section{Counterexample to a claim from~\cite{DBLP:journals/algorithmica/BuchinKLS19}}\label{app:counterex}

The algorithm for computing the Beer-index of a polygon with holes from~\cite{DBLP:journals/algorithmica/BuchinKLS19} works as follows.
We consider a convex polygon $P$ and a family $H_1,\ldots,H_k$ of obstacles, and we want to find the probability that for two random points $p,q$ in $P\setminus \bigcup H_i$, the line segment $pq$ is disjoint from all obstacles.
Note that this is just a slight reformulation of the problem of computing the Beer-index of a polygon with holes.

We partition $P\setminus \bigcup H_i$ into trapezoids $T_1,\ldots,T_t$.
We now compute the Beer-index as the sum of two entities: (i) the probability that two random points $p,q$ are in the same trapezoid, and (ii) the probability that two random points $p,q$ are in different trapezoids and the segment $pq$ is disjoint from all obstacles.
For the second case, we compute the contribution for each pair $T_i,T_j$ of trapezoids and sum over all these pairs.

Let us consider a fixed pair $T_i,T_j$ of trapezoids.
For each pair $s_1^i,s_2^i$ of two segments of $T_i$ and each pair $s_1^j,s_2^j$ of two segments of $T_j$, we do as follows.
Using geometric duality, we map each line intersecting all four segments $s_1^i,s_2^i,s_1^j,s_2^j$ to a point in the dual space, and these points together form a convex polygon $L^*$.
Likewise, the dual points of the set of lines intersecting an obstacle $H_l$ form an ``hourglass-shaped'' unbounded region $H_l^*$ in the dual space.
We now compute the overlay of $L^*$ with all the regions $H_l^*$.
This results in an arrangement $\mathcal S^*$ of the plane.
Each cell $C$ of $\mathcal S^*$ contained in $L^*$ corresponds to a set of lines in primal space that all cross the four segments $s_1^i,s_2^i,s_1^j,s_2^j$.
Furthermore, the visibility from $T_i$ to $T_j$ is either not blocked along any of these lines or blocked for all the lines by one or more obstacles.
It is then described in the paper how to compute the contribution to the Beer-index of all pairs of points $p\in T_1$, $q\in T_2$, where the dual point $l(pq)^*$ is in $C$.
The time it takes to compute this contribution is proportional to the complexity of $C$.

Running over all cells $C$ contained in $L^*$, we obtain that computing the contribution for all pairs of points $p\in T_1$, $q\in T_2$, where $l(pq)$ intersects $s_1^i,s_2^i,s_1^j,s_2^j$, can be done in $O(n^2)$ time, where $n$ is the total complexity of $P$ and the obstacles.
Since there are $O(1)$ choices of the four segments $s_1^i,s_2^i,s_1^j,s_2^j$, the contribution from all pairs $p\in T_1$, $q\in T_2$ can then also be computed in time $O(n^2)$.

\begin{figure}
\centering
\includegraphics[page=7]{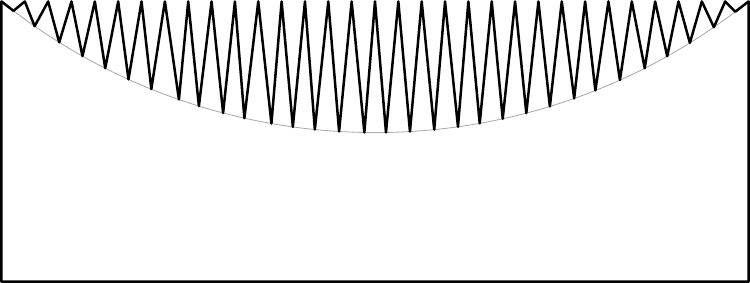}
\caption{For polygons of this type, the algorithm from~\cite{DBLP:journals/algorithmica/BuchinKLS19} has running time $\Omega(n^4)$.}
\label{fig:counterex}
\end{figure}

It is then erroneously claimed in Lemma 3 that the total size of the arrangements $\mathcal S^*$, as we sum over all pairs of trapezoids $T_i,T_j$, is also $O(n^2)$.
This is not true.
For $\Omega(n^2)$ of the pairs, the complexity may be $\Omega(n^2)$, resulting in a total complexity of $\Omega(n^4)$.
This can even be the case for simple polygons, as the polygon in \Cref{fig:counterex} shows.
The polygon is simply a rectangle with some spikes removed along the top edge.
The obstacles will therefore be the removed spikes.
Note that a line can intersect any consecutive interval of obstacles and no other obstacle, so there are $\Omega(n^2)$ subsets of obstacles that can be intersected.
The arrangement we get by overlaying the hour-glasses $H_l^*$ therefore has complexity $\Omega(n^2)$.
It then follows that for $\Omega(n^2)$ pairs of trapezoids $T_i,T_j$, the arrangement $\mathcal S^*$ has complexity $\Omega(n^2)$, resulting in a running time of $\Omega(n^4)$.

\section{Triangle integrals}\label{sec:triangleintegrals}

We now explain how to compute the integral
\[
\Xi(abc)=
\iint_{q\in \Delta(abc)} \left\Vert aq\right\Vert \diff q.
\]
By rotation and scaling, we can assume that $a=(0,0)$ and $y(b)=y(c)=1$, where $b$ is to the left of $c$.
We make a transformation of coordinates so that the usual coordinates corresponding to the pair $(\rho, q)$ is
\[
F(\rho,q)=q\cdot (b\cdot (1-\rho)+c\cdot \rho)=q\cdot (x(b)\cdot (1-\rho) + x(c)\cdot \rho,1).
\]
We then have the Jacobian
\[
J_F =
\begin{bmatrix}
q\cdot (-x(b) + x(c)) & x(b)\cdot (1-\rho) + x(c)\cdot \rho \\
0 & 1
\end{bmatrix},
\]
with the determinant $\det J_F=q\cdot (x(c)-x(b))$.
As $x(b)<x(c)$ and $q\in[0,1]$, we have $\vert \det J_F\vert=q\cdot (x(c)-x(b))$.

We then have
\[
\Xi(abc) = \int_0^1\int_0^1 q\, \sqrt{(x(b)\cdot (1-\rho) + x(c)\cdot \rho)^2 + 1} \left\vert\det J_F\right\vert \diff q\diff \rho.
\]
Using Maple~\cite{maple}, we can find a find a formula for this integral.

\begin{figure}
\centering
\includegraphics[page=21]{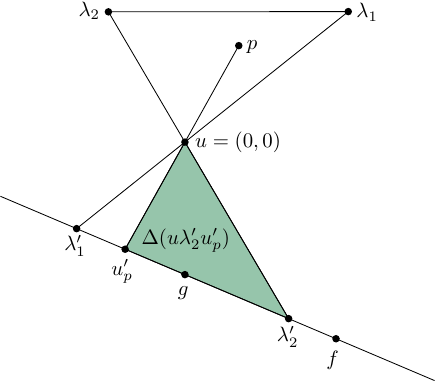}
\caption{The setting of integrals \eqref{eq:triint2} and \eqref{eq:triint3}.}
\label{fig:integralsetting}
\end{figure}

We now turn our attention to the integrals \eqref{eq:triint2} and \eqref{eq:triint3}.
We focus on \eqref{eq:triint2} since \eqref{eq:triint3} is analogous; see \Cref{fig:integralsetting}.
The corners of $\Lambda$ are $u,\lambda_1,\lambda_2$, and by rotation, scaling and translation, we can assume that $u=(0,0)$, $y(\lambda_1)=y(\lambda_2)=1$ and $x(\lambda_2)<x(\lambda_1)$.
Recall that $u\lambda_1'$ and $u\lambda_2'$ are the extensions of $\lambda_1u$ and $\lambda_2u$ from $u$.
Let the edge containing $\lambda_1'$ and $\lambda_2'$ be defined by corners $f$ and $g$.
Unless $fg$ is vertical, we can assume $x(f)=1$ and $x(g)=0$ by sliding the points on the line containing the edge.
We can then express the integral in terms of the four parameters $x(\lambda_1),x(\lambda_2),y(f),y(g)$.

We use the same transformation of coordinates to pairs $(\rho,q)$ as used above for the integral $\Xi(abc)$.
We rewrite \eqref{eq:triint2} as the integral
\[
\int_0^1\int_0^1 q\, \sqrt{(x(\lambda_1)\cdot (1-\rho) + x(\lambda_1)\cdot \rho)^2 + 1} \left\vert \Delta(\rho)\right\vert  \left\vert\det J_F\right\vert \diff q\diff \rho,
\]
where $\Delta(\rho)$ is shorthand for the triangle corresponding to $\Delta(u\lambda'_2u'_p)$ (note that using the transformed coordinates $(\rho,q)$, the triangle depends only on $\rho$).

As it turns out, Maple refuses to evaluate this integral directly, but we can evaluate the inner integral and we observe that the result has the form
\[
\frac{(a\,\rho + b)\, \sqrt{c\,\rho^2+d\, \rho+e}}{f\, \rho+g},
\]
where $a,b,c,d,e,f$ are constants depending on the parameters $x(\lambda_1),x(\lambda_2),y(f),y(g)$.
Maple is able to find an anti-derivative of this function.
By inserting the expressions for $a,b,c,d,e,f$, we can then evaluate the integral \eqref{eq:triint2}.

\newpage

\section{Maple calculation}\label{app:maple}

The following code shows a calculation of a closed-form expression for the integral~\eqref{eq:int4}.

\includegraphics[page=1,width=.95 \textwidth]{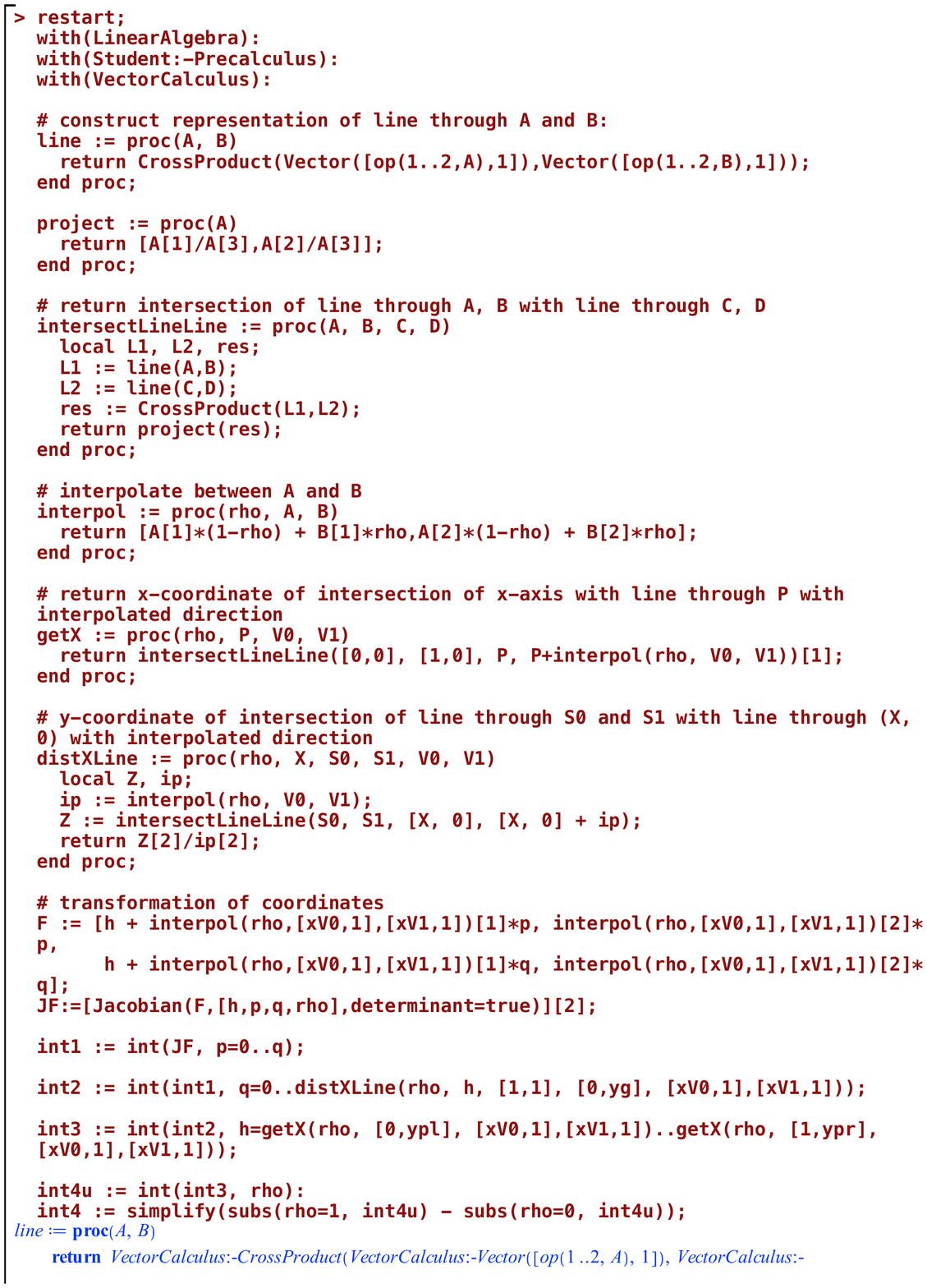}

\includegraphics[page=2,width=.95 \textwidth]{trapezoidIntegral.pdf}

\includegraphics[page=3,width=.95 \textwidth]{trapezoidIntegral.pdf}

\end{document}